\documentclass[12pt]{article}
\usepackage[utf8]{inputenc}
\usepackage[T1]{fontenc}
\usepackage[top=2.5cm, left=2cm, right=2cm, bottom=2.35cm]{geometry}
\usepackage{pstricks,pst-node,pst-text,pst-3d}

\usepackage{soul}
\usepackage{amsthm}
\usepackage{algorithm}
\usepackage{algorithmicx}
\usepackage[noend]{algpseudocode}
\usepackage{amsmath}
\usepackage{amssymb}
\usepackage[mathscr]{eucal}
\usepackage{bigfoot}
\usepackage{bbm}
\usepackage{xcolor}
\usepackage{mathtools}
\usepackage{hyperref}
\usepackage{bm}
\usepackage{enumerate}
\usepackage{booktabs}
\usepackage{authblk}

\begin{document}

\newtheorem{theorem}{Theorem}[section]
\newtheorem{proposition}[theorem]{Proposition}
\newtheorem{corollary}[theorem]{Corollary}
\newtheorem{lemma}[theorem]{Lemma}
\newtheorem{conjecture}[theorem]{Conjecture}
\theoremstyle{definition}
\newtheorem{example}[theorem]{Example}
\newtheorem{hypotheses}[theorem]{Hypothèses}
\newtheorem{definition}[theorem]{Definition} 
\newtheorem{notation}[theorem]{Notation}
\newtheorem{commentaires}[theorem]{Commentaires}
\newtheorem{remark}[theorem]{\textit{Remark}}

\newcommand{\bbT}{\mathbb{T}}
\newcommand{\bbR}{\mathbb{R}}
\newcommand{\bbH}{\mathbb{H}}
\newcommand{\bbN}{\mathbb{N}}
\newcommand{\bbZ}{\mathbb{Z}} 
\newcommand{\bbC}{\mathbb{C}}
\newcommand{\bbE}{\mathbb{E}}
\newcommand{\bbB}{\mathbb{B}}
\newcommand{\bbS}{\mathbb{S}}
\newcommand{\bbI}{\mathbb{I}}
\newcommand{\bfone}{\mathbf{1}}
\newcommand{\calI}{\mathscr{I}}
\newcommand{\br}{{\rm br}}
\newcommand{\BR}{{\rm BR}}
\newcommand{\core}{{\rm core}}
\newcommand{\calG}{\mathscr{G}}
\newcommand{\calA}{\mathscr{A}}
\newcommand{\calD}{\mathscr{D}}
\newcommand{\calH}{\mathscr{H}}
\newcommand{\calN}{\mathscr{N}}
\newcommand{\calZ}{\mathscr{Z}}
\newcommand{\calS}{\mathscr{S}}
\newcommand{\calP}{\mathscr{P}}
\newcommand{\calF}{\mathscr{F}}
\newcommand{\calC}{\mathscr{C}}
\newcommand{\calB}{\mathscr{B}}
\newcommand{\calE}{\mathscr{E}}
\newcommand{\calU}{\mathscr{U}}
\newcommand{\calL}{\mathscr{L}}
\newcommand{\calV}{\mathscr{V}}
\newcommand{\calFC}{\mathscr{FC}}
\newcommand{\calVE}{\mathscr{VE}}
\newcommand{\Hom}{{\rm Hom}}
\newcommand{\domS}{\ {\rm dom}_S \ }
\newcommand{\domP}{\ {\rm dom}_P \ }
\newcommand{\dom}{\ {\rm dom} \ }
\newcommand{\tto}{\mathtt{0}}
\newcommand{\ttl}{\mathtt{1}}
\newcommand{\bbone}{\mathbbm{1}}
\newcommand{\ext}{{\rm ext}}
\newcommand{\rk}{{\rm rk}}
\newcommand{\Span}{{\rm span}}
\newcommand{\Ima}{{\rm Im}}
\newcommand{\Ker}{{\rm Ker}}

\newrgbcolor{purple}{0.5 0 .5}

\makeatletter
\newenvironment{breakablealgorithm}
{ % \begin{breakablealgorithm}
   \begin{center}
     \refstepcounter{algorithm}% New algorithm
     \rule{\linewidth}{1pt} %height.8pt depth0pt \kern2pt% \@fs@pre for \@fs@ruled
     \renewcommand{\caption}[2][\relax]{% Make a new \caption
       {\raggedright\textbf{\ALG@name~\thealgorithm} ##2\par}%
       \ifx\relax##1\relax % #1 is \relax
         \addcontentsline{loa}{algorithm}{\protect\numberline{\thealgorithm}##2}%
       \else % #1 is not \relax
         \addcontentsline{loa}{algorithm}{\protect\numberline{\thealgorithm}##1}%
       \fi
       \vspace{-9pt}
      \rule{\linewidth}{0.5pt} %\kern2pt\hrule\kern2pt
      \vspace{-15pt}
     }
  }{% \end{breakablealgorithm}
     \vspace{-9pt}
     \rule{\linewidth}{0.5pt} %\kern2pt\hrule\relax% \@fs@post for \@fs@ruled
   \end{center}
 }
\makeatother

\providecommand{\keywords}[1]
{
  \small	
  \textbf{Keywords---} #1
}

\providecommand{\JEL}[1]
{
  \small	
  \textbf{\noindent\textit{JEL Classification:}} #1
}

\providecommand{\MSC}[1]
{
  \small	
  \textbf{\noindent\textit{MSC Subject Classification:}} #1
}

\renewcommand{\Return}{\State \textbf{return} }

\author{Dylan Laplace Mermoud\thanks{Centre d'Économie de la Sorbonne, Université Paris 1 Panthéon-Sorbonne.},\thanks{Corresponding author (dylan.laplace.mermoud@gmail.com)} \ Michel Grabisch\thanks{Paris School of Economics, Université Paris 1 Panthéon-Sorbonne.}, and Peter Sudhölter\thanks{Department of Economics, University of Southern Denmark.}}

\title{Minimal balanced collections and their application to core stability and other topics of game theory\footnote{The authors are grateful for valuable remarks of a referee and the area editor who helped to improve this article.}}

\date{}%\date{Version of \today}

\maketitle

\abstract{Minimal balanced collections are a generalization of partitions of a finite set of $n$ elements and have important applications in cooperative game theory and discrete mathematics. However, their number is not known beyond $n=4$. In this paper we investigate the problem of generating minimal balanced collections and implement the Peleg algorithm, permitting to generate all minimal balanced collections till $n=7$. Secondly, we provide practical algorithms to check many properties of coalitions and games, based on minimal balanced collections,  in a way which is faster than linear programming-based methods. In particular, we construct an algorithm to check if the core of a cooperative game is a stable set in the sense of von Neumann and Morgenstern. The algorithm implements a theorem according to which the core is a stable set if and only if a certain nested balancedness condition is valid. The second level of this condition requires generalizing the notion of balanced collection to balanced sets.}

\medskip 



\section{Introduction}\label{sec:intro}
The term ``balanced collection'' has been coined by Shapley in his seminal paper
\cite{sha67}, giving necessary and sufficient conditions for the core of a
cooperative game to be nonempty. Cooperative game theory aims at defining
rational ways (called solutions of a game) in sharing a benefit obtained by
cooperation of a set of players. The core of a game (Gillies \cite{gil59}) is
one of the most popular concepts of solution of a game, and appears in other
domains like decision theory and combinatorial optimization.

A balanced collection is a collection of subsets of a finite set $N$, and can be
seen as a generalization of partitions of $N$, in the sense that weights are
assigned to each subset in the collection, in order that each element of $N$
receives a total weight equal to 1. In fact, mainly minimal balanced collections
are of interest, that is, balanced collections for which no proper subcollection
is balanced.  Apart from their use in cooperative game theory,
balanced collections appear in several domains of discrete mathematics (e.g.,
hypergraphs), combinatorics (e.g., combinatorial design theory), while their
counterpart, namely, unbalanced collections (more precisely, maximal unbalanced
collections), appear in quantum physics.

Being a generalization of partitions, (minimal) balanced collections are highly
combinatorial objects, and their number is so far not known beyond
$|N|=4$. However, in the domain of cooperative game theory, Peleg \cite{pel65}
has proposed a recursive algorithm to generate them, which, as far as we know,
has never been implemented.

\medskip

The first achievement of this paper is to provide algorithms generating minimal balanced collections. We begin by implementing Peleg's algorithm and generating all minimal balanced collections till $n=7$. We
  show also that minimal balanced collections can be generated via a vertex
  enumeration method, hereby relating the problem of generating minimal balanced
  collections to one of the fundamental open problems in geometry
  \cite{boelguma09}. Applying Avis and Fukuda's method for vertex enumeration \cite{avfu92},
  we find the former method by Peleg more efficient.

The second achievement of the paper is to show that minimal
  balanced collections are a central concept in cooperative game theory that can
  be applied to check a variety of properties, and most importantly, core
  stability. The question to check whether the core of a game is a stable set in
  the sense of von Neumann and Morgenstern has remained an open problem for a
  long time, and was recently solved by Grabisch and Sudh\"olter \cite{grsu20}. It
  turns out that the test of core stability amounts to a complex nested
  balancedness condition, which needs in particular to identify exact
  coalitions, strictly vital exact coalitions, extendable coalitions, and
  feasible collections of coalitions. We show that all these notions can be
  tested or generated via the use of minimal balanced collections, and that this
  way is faster than linear programming methods. The main reason is that minimal
  balanced collections do not depend on the game considered but only on the
  number of players. As a consequence, minimal balanced collections need to be
  generated only once, and can be used repeatedly for any game. We have
  implemented all these algorithms as computer programs, that allow to
  obtain finally a general algorithm to check core stability.

The third achievement is a direct consequence of the nested balancedness
condition of core stability. It happens that this condition requires a
more general notion of balanced collection, which we call (minimal) balanced
sets. A finite set in the nonnegative orthant is balanced if the
  characteristic vector of $N$ is a positive linear combination of its
  elements. We show that this generalized notion causes difficulties and does
not seem to be easily generated by methods applied for minimal balanced
collections. Again, this problem can be related to a vertex
  enumeration problem.

\medskip

The paper is organized as follows. Section~\ref{sec:back} introduces cooperative
games, stable sets and balanced collections. Section~\ref{sec:bcother} shows
where minimal balanced collections appear in different domains and introduces
unbalanced collections, which are used in quantum
physics. Section~\ref{sec:gener} focuses on the generation of minimal balanced
collections, while Section~\ref{sec:appl} describes
applications in cooperative game theory in detail. Section~\ref{sec:mbs} introduces
minimal balanced sets, which is a key notion in the nested balancedness
condition for testing core stability, the object of
Section~\ref{sec:cost}. Section~\ref{sec:conc} concludes the paper. Details on
the use of maximal unbalanced collections in quantum physics are given in
Appendix~\ref{app:A}, while Appendix~\ref{app:B} illustrates by examples the
Peleg algorithm. Finally, all algorithms presented in the paper are given in
pseudo-code, mostly in Appendix~\ref{app:C}, or in the main text.

\section{Background on TU-games and balanced collections} \label{sec:back}
\subsection{TU-games}\label{sec:TU}
Let $N$ be a finite nonempty set of $n$ players and let $2^N$ denote its power
set, i.e., the set of all subsets of $N$. In most cases, we tacitly assume that
$N = \{1, \ldots, n\}$. A nonempty subset of $N$ is called a \emph{coalition}. A
\emph{(cooperative TU) game} is a pair $(N,v)$ with $v: 2^N \to \bbR$ such that
$v(\varnothing) = 0$.

\medskip

Let $(N,v)$ be a game. An \emph{allocation} or {\it payoff vector} is an
$n$-dimensional vector $x = (x_i)_{i \in N}\in\bbR^N$, representing the
distribution of payoffs among the players. Denote by $x(S) = \sum_{i \in S} x_i$ the
payoff received by coalition $S \subseteq N$ with allocation $x$.

An allocation is said to be \emph{efficient} for the game $(N,v)$ if $x(N) = v(N)$, and an efficient
allocation is called a \emph{preimputation}. The set of preimputations is
denoted by $X(N,v)$.

An allocation $x$ is \emph{individually rational} for the game $(N,v)$ if $x_i \geq v(\{i\})$ for
every player $i \in N$. An individually rational preimputation is called an
\emph{imputation}, and the set of imputations is denoted by $I(N,v)$.

An allocation $x$ is {\it coalitionally rational} for the game $(N,v)$ if $x(S)\geq v(S)$ for every
coalition $S\in 2^N$. The {\it core}  \cite{gil53,gil59} of the game $(N,v)$ is the set of
coalitionally rational preimputations, i.e.,
\[
C(N,v) = \{x \in \bbR^N \mid x(S) \geq v(S), \forall S \subseteq N, x(N)=v(N)\}.
\]

\subsection{Stable sets}\label{sec:stab}
A preimputation $x \in X(N,v)$ \emph{dominates via a coalition} $S \subseteq N$
another preimputation $y$ if $x(S) \leq v(S)$ and $x_i > y_i$ for every $i \in
S$, written $x \domS y$. If there exists a coalition $S$ such that $x \domS y$,
then $x$ \emph{dominates} $y$, which is denoted by $x \dom y$.

Based on this notion, von Neumann and Morgenstern \cite{vnm44} introduced the concept of
stable sets for cooperative games. A set $U \subseteq I(N,v)$ is a \emph{stable
  set} if it satisfies
\begin{enumerate}
\item[(i)] \emph{Internal stability}: if $y \in U$ is dominated by $x \in I(N,v)$, then $x \not \in U$,
\item[(ii)] \emph{External stability}: $\forall y \in I(N,v) \setminus U, \exists x \in U$ such that $x \dom y$.
\end{enumerate}

Due to their important stability properties, von Neumann and Morgenstern regard stable sets as the main solution concept for cooperative games and call them, hence, {\it solutions}.  However, although
intuitively appealing, considering stable sets is problematic. Indeed, they may be not
unique, and there exist games without stable sets (see Lucas
\cite{luc71}). Moreover, they are in general difficult to identify. According to
Deng and Papadimitriou \cite{depa94}, the existence of a stable set may be
undecidable. These difficulties have led to the development of other solution
concepts, among which the core (Gillies \cite{gil59}, see above) is the most popular. The
computation of the core is relatively easy but expensive, due to the large
number of inequalities defining it, and it is empty for large classes of games. 

Both stable sets and the core have their own merits as solution
concepts. Indeed, the notions of domination and stability are highly intuitive,
``coalitional rationality'' is a desirable property, and its easy computability
supports the core. By definition, the core is contained in each stable
set. Hence, if the core is (externally) stable, it must be the unique stable
set. Therefore, it is an interesting and important problem to characterize the
set of games for which the above-mentioned solution concepts coincide, i.e., to
provide necessary and sufficient conditions for external stability of the
core. This is what Grabisch and Sudh\"olter have achieved \cite{grsu20}. We will
come back to this result in Section~\ref{sec:cost}.

\subsection{Balanced collections}\label{sec:bc}
We use throughout the paper the following notation: For any $T \subseteq N$, its
{\it characteristic vector} $\bfone^T\in\bbR^N$ is defined by $\bfone^T_i = 1$ if $i \in
T$, and $\bfone^T_i = 0$ otherwise.

\begin{definition}
A collection $\calB$ of coalitions in $2^N$ is \emph{balanced} if there exists a system of positive weights $(\lambda_S)_{S \in \calB}$, called \emph{balancing weights}, such that $\sum_{S \in \calB} \lambda_S \bfone^S = \bfone^N$. 
\end{definition}
Therefore, $\calB$ is balanced if and only if the vector
$\bfone^N$ is in the relative interior of the cone
generated by the vectors $\bfone^S, S\in\calB$.

Let us give some examples of balanced collections:
\begin{enumerate}
\item Every partition of $N$ is a balanced collection, where every set belonging
  to the partition has balancing weight equal to 1;
\item For $n=3$: $\{\{1,2\},\{1,3\},\{2,3\}\}$ is balanced with
  weights $\big(\frac{1}{2},\frac{1}{2},\frac{1}{2}\big)$.  More generally,
    every anti-partition (the set of complements of a partition of $s\geq 2$
    blocks) is a balanced collection, where the balancing weight of each element
    is $\frac{1}{s-1}$;
\item For $n=4$: $\{\{1,2\},\{1,3\},\{1,4\},\{2,3,4\}\}$ is balanced with
  weights $\big(\frac{1}{3},\frac{1}{3},\frac{1}{3},\frac{2}{3}\big)$.
\end{enumerate}

A balanced collection is \emph{minimal} if it does not contain a balanced proper
subcollection. We denote by $\bbB(N)$ the set of minimal balanced collections on
$N$. It is well known that a balanced collection is minimal if and only if its
system of balancing weights is unique. Note that all the examples above are minimal
balanced collections.

\medskip

As the subsequent sections will show, minimal balanced collections are highly
combinatorial objects, much more than partitions, and they appear in many
domains. Their number for a given $n$
is not known beyond $n=4$. Although Peleg \cite{pel65} proposed an inductive
algorithm to generate them, up to our knowledge it has never been implemented,
so that there is no available list of minimal balanced collections for $n=5$ and
larger. Thanks to our own implementation of Peleg's algorithm, we are able
  to give the list of minimal balanced collections till $n=7$ (see
  Section~\ref{sec:Pele}).

\section{Balanced collections in other domains}\label{sec:bcother}
Balanced collections are known in other fields of discrete mathematics under
different names, especially in hypergraphs and combinatorial design theory. We
start with hypergraphs.

\subsection{Hypergraphs} 
(see Berge \cite{ber89}) An \emph{(undirected) hypergraph} $\calH$ is a pair
$\calH = (X, E)$ where $X$ is the set of elements called \emph{nodes} or
\emph{vertices}, and $E$ is a  collection of non-empty subsets of $X$,
called \emph{hyperedges} or \emph{edges}.  A hypergraph is {\it simple} if
  there is no repetition in the collection of hyperedges. 
To get rid of repetitions, it is convenient to consider any hypergraph as a simple hypergraph $(X,E)$ to
which we assign a weight (or multiplicity) function  $w:E \to \bbN$ on hyperedges, defining the
degree of a vertex $x$ by
\[
d^w_\calH(x) = \sum_{e \in \calH(x)} w(e).
\]

We denote by $\calH(x)$ the family of
hyperedges $e\in E$ containing the vertex $x$.

\begin{definition}
A hypergraph $(X,E)$ is \emph{$d$-regular} if every vertex has \emph{degree}
$d$.  A hypergraph is {\it
    regular} if it is $d$-regular for some $d$.
\end{definition}

\medskip

It is easy to see that regular hypergraphs are in bijection with
balanced collections. Indeed, consider a regular hypergraph
$\calH=(X,E)$ with weight (multiplicity) function $w$. Denote by $\delta
\coloneqq d^w_\calH(x)$ the degree of any vertex $x \in X$, and define the
weight system $(\lambda_e)_{e\in E}$ by
\[
\lambda_e \coloneqq \frac{w(e)}{\delta}, \quad \forall e\in E.
\]
Then, $E$, wiewed as a set of subsets of $X$, is a balanced collection on
$X$ with balancing weight system  $(\lambda_e)_{e\in E}$. Conversely, any
balanced collection on $X$ can be seen as a regular hypergraph.

In \cite{sha67}, Shapley calls the value $\delta$ the \emph{depth} of the associated balanced collection, and $w(S)$ the \emph{multiplicity} of the coalition $S$ in the balanced collection.

\medskip

There is another concept in hypergraph theory which corresponds to balanced
collections.
\begin{definition}
A \emph{fractional matching} in a hypergraph $\calH=(X,E)$ is a function $\mu: E \to [0,1]$ such that for every vertex $x$ in $X$,
\[
\sum_{e \in \calH(x)} \mu(e) \leq 1.
\] 
A fractional matching is called \emph{perfect} if for every vertex $x$ in $X$, 
\[
\sum_{e \in \calH(x)} \mu(e) = 1. 
\] 
\end{definition}
It is easy to see that any simple hypergraph on $X$ admitting a perfect
fractional matching $\mu$ induces a balanced collection on $X$, which is $\{e\in
E\mid \mu(e)>0\}$. The converse holds as well.

\subsection{Combinatorial design theory} It is the part of combinatorial
mathematics that deals with the existence, construction, and properties of
systems of finite sets whose arrangements satisfy generalized concepts of
\emph{balance} and/or \emph{symmetry}. The following definitions come from
Colbourn and Dinitz \cite{codi06}.

\begin{definition}
A \emph{$t$-wise balanced design} of type $t$-$(n, K, \delta)$ is a pair $(N,
\calB)$ where $N$ is a set of $n$ elements, called \emph{points}, and $\calB$ is
a collection of subsets of $N$, called \emph{blocks}, with the property that the
size of every block belongs to the set $K$, and every subset of size $t$ of $N$
is contained in exactly $\delta$ blocks.
\end{definition}
In the context of combinatorial design, the value $\delta$ is called the
\emph{replication number}. With a slightly different notation, $t$-wise balanced
designs are also called {\it $t$-balanced incidence structures}.

By a mechanism similar to the one used with hypergraphs, it is easy to see that
1-wise balanced designs are balanced collections, and the converse is also true
provided repetition is allowed in $\calB$.

\subsection{Unbalanced collections and hyperplanes arrangements}
A collection of subsets of $N$ which is not balanced is said to be {\it
  unbalanced}. It is {\it maximal} if no supercollection of it is unbalanced.
Strangely enough, maximal unbalanced collections are also an important
topic of discrete mathematics, with applications in physics (see Billera et
al. \cite{bimomowawi12}). 

As for minimal balanced collections, there is no closed-form formula to compute
the number of maximal unbalanced collections. Their number is known till $n=9$
(see Table~\ref{tab:muc}).
\begin{table}
  \begin{center}
    \begin{tabular}{|c|r|}\hline
    $n$ &Nb of maximal unbalanced collections\\ \hline
    2 & 2 \\
    3 & 6 \\
    4 & 32 \\
    5 & 370\\
    6 & 11,292\\
    7 &1,066,044\\
    8 &347,326,352\\
    9 & 419,172,756,930\\ \hline
    \end{tabular}
  \end{center}
    \caption{Number of maximal unbalanced collections as a function of $n$}
  \label{tab:muc}
\end{table}
In order to characterize unbalanced collections, we recall a result from Derks
and Peters \cite{depe98}.
\begin{proposition}
A collection $\calS \subseteq 2^N$ of nonempty sets is balanced if and only if for every vector $y \in \bbR^N$ such that $\sum_{i \in N} y_i = 0$, either $\sum_{i \in S} y_i = 0$ for every $S \in \calS$ or there exist $S, T \in \calS$ such that $\sum_{i \in S} y_i > 0$ and $\sum_{i \in T} y_i < 0$. 
\end{proposition}
Therefore, a collection $\calS$ of nonempty sets is unbalanced if and only if there exists
$y \in \bbR^N$ such that $\sum_{i \in N} y_i = 0$ and $\sum_{i \in S} y_i > 0$
for all $S \in \calS$. Let us give two examples of maximal unbalanced
collections with a possible vector $y$:
\begin{itemize}
\item[(i)] For $n=3$: $\{\{1,2\},\{1,3\},\{1\}\}$, $y=(2, -1, -1)$;
\item[(ii)] For $n=4$:
  $\{\{1\},\{1,2\},\{1,3\},\{1,4\},\{1,2,3\},\{1,2,4\},\{1,3,4\}\}$, $y=(3,-1,-1,-1)$.
\end{itemize}

\begin{figure}[htb]
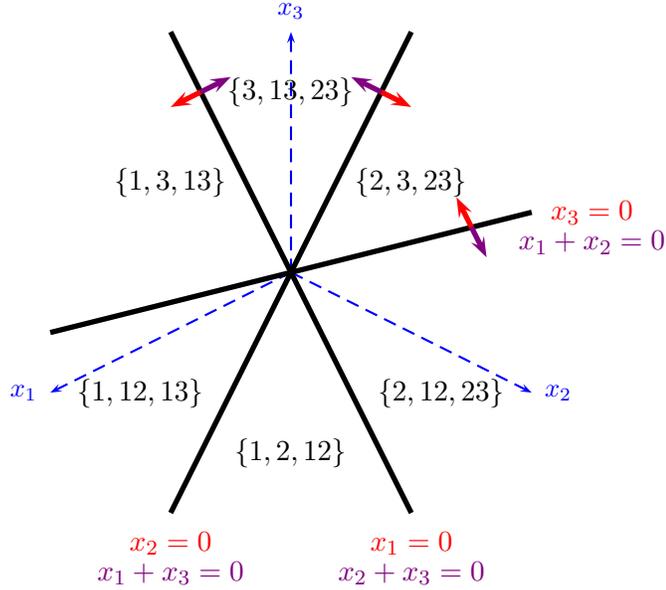

  \begin{center}
  \psset{unit=0.8cm}
  \pspicture(-5,-5)(5,5)
  \psline[linestyle=dashed,linecolor=blue]{->}(0,0)(0,4)
  \psline[linestyle=dashed,linecolor=blue]{->}(0,0)(4,-2)
  \psline[linestyle=dashed,linecolor=blue]{->}(0,0)(-4,-2)
  \uput[180](-4,-2){\blue\footnotesize $x_1$}
  \uput[0](4,-2){\blue\footnotesize $x_2$}
  \uput[90](0,4){\blue\footnotesize $x_3$}
  \psline[linewidth=2pt](-2,4)(2,-4)
  \psline[linewidth=2pt](2,4)(-2,-4)
  \psline[linewidth=2pt](-4,-1)(4,1)
  \psline[linewidth=2pt,linecolor=red]{->}(1.5,3)(2,2.75)
  \psline[linewidth=2pt,linecolor=purple]{->}(1.5,3)(1,3.25)
  \psline[linewidth=2pt,linecolor=red]{->}(-1.5,3)(-2,2.75)
  \psline[linewidth=2pt,linecolor=purple]{->}(-1.5,3)(-1,3.25)
  \psline[linewidth=2pt,linecolor=red]{->}(3,0.75)(2.75,1.25)
  \psline[linewidth=2pt,linecolor=purple]{->}(3,0.75)(3.25,0.25)
  \rput(-2,-4.5){\small \red $x_2=0$}
  \rput(-2,-5){\small \purple $x_1+x_3=0$}
  \rput(2,-4.5){\small \red $x_1=0$}
  \rput(2,-5){\small \purple $x_2+x_3=0$}
  \rput(5,1){\small \red $x_3=0$}
  \rput(5,0.5){\small \purple $x_1+x_2=0$}
  \rput(0,-3){\small $\{1,2,12\}$}
  \rput(2.5,-2){\small $\{2,12,23\}$}
  \rput(2,1.5){\small $\{2,3,23\}$}
  \rput(0,3){\small $\{3,13,23\}$}
  \rput(-2,1.5){\small $\{1,3,13\}$}
  \rput(-2.5,-2){\small $\{1,12,13\}$}
  \endpspicture
  \end{center}
  \caption{The restricted all-subset arrangement for $n=3$ in the plane
    $H_N$. Arrows indicate the normal vector to the hyperplane of the same
    color. The 6 maximal unbalanced collections (subsets are written without
    comma and braces) correspond to the 6 regions.}
  \label{fig:hyp}
\end{figure}

This very simple characterization permits to see that maximal unbalanced
collections are the same as Bj\"orner’s positive set sum systems \cite{bjo15}, and that they are related to hyperplanes arrangements. In
the hyperplane $H_N=\{x\in\bbR^N\mid x(N)=0\}$, consider the hyperplanes $\{x\in
H_N\mid x(S)=0\}$, for all $S\in 2^N\setminus\{\varnothing,N\}$. Observe that
since we are in $H_N$, the hyperplanes induced by $S$ and $N\setminus S$ are the
same, but their normal vectors point in opposite directions. It follows that
these $2^{n-1}-1$ distinct hyperplanes (called by Billera et al. {\it restricted
  all-subset arrangement}) define full-dimensional elementary regions $R_\calS$,
each one characterized by the collection $\calS=\{S\in
2^N\setminus\{\varnothing,N\}\mid x(S)>0\}$ where $x$ is any element of
$R_\calS$, and with the property
$|\calS|=2^{n-1}-1$. In addition, each elementary region $R_\calS$ corresponds
to a maximal unbalanced collection $\calS$, and conversely. This shows that each
maximal unbalanced collection has $2^{n-1}-1$ sets, and their number is the
number of elementary regions induced by the hyperplane arrangement (see Figure~\ref{fig:hyp}).

As remarked by Billera et al., the above hyperplane arrangement (and consequently,
maximal unbalanced collections) appears in the field of thermal quantum
physics. We give some details on this application in Appendix~\ref{app:A}.

\section{Generation of minimal balanced collections}\label{sec:gener}
So far, there are two known methods for generating minimal balanced
collections. The first one, due to Peleg \cite{pel65}, is specifically devoted
to the generation of minimal balanced collections and proceeds by
induction on the number of players $n$. The second one uses any vertex
enumeration method for convex polyhedra, applied on a specific polytope whose
vertices correspond to the minimal balanced collections. 

\subsection{Peleg's algorithm}\label{sec:Pele}
Peleg \cite{pel65} developed an inductive method to construct, from the minimal
balanced collections defined on a set $N$, all those that are defined on the set
$N' = N \cup \{p\}$, with $p$ a new player that was not included in $N$. As far as we know, Peleg's inductive method has never been
implemented as an algorithm, perhaps due to the rather abstract way it is described, far from
any algorithmic considerations. For this reason, we translate Peleg's method and
results from an algorihmic point of view, reproving his results in our new
formalism for the sake of clarity and completeness. In the following, the main
result is divided into four cases and the fourth one is slightly generalized.

\medskip

Let $\calC = \{S_1, \ldots, S_k\}$ be a balanced collection of $k$ coalitions on
$N$. Denote by $[k]$ the set $\{1, \ldots, k\}$ for any positive integer $k$. If
$\lambda^\calC$ is a system of balancing weights for $\calC$ and $I \subseteq
[k]$ is a subset of indices, denote by $\lambda^\calC_I$ the sum $\sum_{i \in I}
\lambda^\calC_{S_i}$. Also, denote by $A^\calC$ the $(n \times k)$-matrix formed
by the $k$ column vectors $\bfone^{S_1},\ldots,\bfone^{S_k}$. Denote by $\rk(A^\calC)$ the rank of the matrix $A^\calC$, meaning the dimension of the
Euclidean space spanned by its columns viewed as $k$-dimensional vectors.

\paragraph{First case.} Assume that $\calC$ is a minimal balanced collection on $N$. Take $I \subseteq [k]$ such that $\lambda^\calC_I = 1$. Denote by $\calC'$ the new collection in which the coalitions $\{S_i\}_{i \in I}$ contain the new player $p$ as additional member and the other coalitions $\{S_j\}_{j \in [k] \setminus I}$ are kept unchanged. 

\begin{lemma}
$\calC'$ is a minimal balanced collection on $N'$. 
\end{lemma}

\proof
Because $\calC$ is a minimal balanced collection, the equalities $\sum_{S \in \calC', S \ni i} \lambda^\calC_S = 1$ are already satisfied for any player $i \in N$. By definition of $I$, we also have that $\sum_{S \in \calC', S \ni p} \lambda^\calC_S = 1$. Then $\calC'$ is balanced. Because $\calC$ is minimal, so is $\calC'$. 
\endproof

\paragraph{Second case.} We assume that $\calC$ is a minimal balanced collection on $N$. Take $I \subseteq [k]$ such that $\lambda^\calC_I < 1$. We denote by $\calC'$ the new collection in which the coalitions $\{S_i\}_{i \in I}$ contain the new player $p$ as additional member, the other coalitions $\{S_j\}_{j \in [k] \setminus I}$ are kept unchanged, and in which the coalition $\{p\}$ is added:
\[
\calC' = \{S_i \cup \{p\} \mid i \in I\} \cup \{S_i \mid i \in [k] \setminus I\} \cup \{\{p\}\}.
\]

\begin{lemma}
$\calC'$ is a minimal balanced collection on $N'$. 
\end{lemma}

\proof
Because $\calC$ is a minimal balanced collection, the equalities $\sum_{S \in \calC', S \ni i} \lambda^\calC_S = 1$ are already satisfied for any player $i \in N$. Define $\lambda^{\calC'}$ such that $\lambda^{\calC'}_S = \lambda^\calC_S$ for $S \in \calC$ and $\lambda^{\calC'}_{\{p\}} = 1 - \lambda^\calC_I$. Therefore
\[
\sum_{\substack{S \in \calC' \\ S \ni p}} \lambda^{\calC'}_S = \lambda^{\calC'}_{\{p\}} + \sum_{i \in I} \lambda^\calC_{S_i} = 1 - \sum_{i \in I} \lambda^\calC_{S_i} + \sum_{i \in I} \lambda^\calC_{S_i} = 1.
\]
Then $\calC'$ is balanced. We cannot obtain a balanced subcollection of $\calC'$ by discarding one of the $\{S_i\}_{i \in [k]}$ because $\calC$ is minimal, and we can also not either discard coalition $\{p\}$ because $\lambda^\calC_I < 1$ and the balancing weights for $\calC$ are unique. So $\calC'$ is minimal. 
\endproof

\paragraph{Third case.} We assume that $\calC$ is a minimal balanced collection on $N$. Take a subset $I \subseteq [k]$ and an index $\delta \in [k] \setminus I$ such that $1 > \lambda^\calC_I > 1 - \lambda^\calC_{S_\delta}$. We denote by $\calC'$ the new collection in which the coalitions $\{S_i\}_{i \in I}$ contain the new player $p$ as additional member, the other coalitions $\{S_j\}_{j \in [k] \setminus I}$ are kept unchanged, and in which the coalition $S_\delta \cup \{p\}$ is added:
\[
\calC' = \{S_i \cup \{p\} \mid i \in I\} \cup \{S_i \mid i \in [k] \setminus I\} \cup \{S_\delta \cup \{p\}\}.
\]

\begin{lemma}
$\calC'$ is a minimal balanced collection on $N'$. 
\end{lemma}

\proof Define $\lambda^{\calC'}$ by $\lambda^{\calC'}_S = \lambda^\calC_S$ for $S \in \calC \setminus \{S_\delta\}$, 
\[
\lambda^{\calC'}_{S_\delta \cup \{p\}} = 1 - \lambda^\calC_I \text{ and } \lambda^{\calC'}_{S_\delta} = \lambda^\calC_{S_\delta} - \lambda^{\calC'}_{S_\delta \cup \{p\}}.
\] 
Let $i \in N$ be a player. If $i \not \in S_\delta$, by balancedness of $\calC$, $\sum_{S \in \calC', S \ni i} \lambda^{\calC'}_S = 1$. If $i \in S_\delta$, then
\[
\sum_{\substack{S \in \calC' \\ S \ni i}} \lambda^{\calC'}_S = \lambda^{\calC'}_{S_\delta \cup \{p\}} + \lambda^{\calC'}_{S_\delta} + \sum_{\substack{S \in \calC \setminus \{S_\delta\} \\ S \ni i}} \lambda^{\calC'}_S = \lambda^\calC_{S_\delta} + \sum_{\substack{S \in \calC \setminus \{S_\delta\} \\ S \ni i}} \lambda^\calC_S = \sum_{\substack{S \in \calC \\ S \ni i}} \lambda^\calC_S,
\]
that is equal to $1$ by balancedness of $\calC$. Concerning player $p$,
\[
\sum_{\substack{S \in \calC' \\ S \ni p}} \lambda^{\calC'}_S = \lambda^{\calC'}_{S_\delta \cup \{p\}} + \lambda^{\calC'}_I = 1 - \lambda^\calC_I + \lambda^\calC_I = 1.
\]
Then $\calC'$ is balanced. Because none of the coalitions $S \in \calC$ or $S_\delta \cup \{p\}$ can be discarded to obtain a balanced subcollection, the proof is finished. 
\endproof

\paragraph{Last case.} In this case, assume that $\calC$ is the union of two different minimal balanced collections on $N$, $\calC^1$, and $\calC^2$, such that the rank of $A^\calC$ is $k-1$. Define two systems of balancing weights for $\calC$, by
\[
\mu_S = \left\{ \begin{array}{l}
\lambda^{\calC^1}_S \text{ if } S \in \calC^1, \\
0 \text{ otherwise}.
\end{array} \right. \quad \quad \quad
\nu_S = \left\{ \begin{array}{l}
\lambda^{\calC^2}_S \text{ if } S \in \calC^2, \\
0 \text{ otherwise}.
\end{array} \right.
\]
Take a subset $I \subseteq [k]$ such that $\mu_I \neq \nu_I$ and 
\[
t^I = \frac{1 - \mu_I}{\nu_I - \mu_I} \in \ ]0, 1[.
\]
Denote by $\calC'$ the new collection in which the coalitions $\{S_i\}_{i \in I}$ contain the new player $p$ as additional member and the other coalition $\{S_j\}_{j \in [k] \setminus I}$ are kept unchanged.

\begin{lemma}
$\calC'$ is a minimal balanced collection on $N'$. 
\end{lemma}

\proof Define $\lambda = (\lambda_S)_{S \in \calC'}$ by $\lambda_S = (1-t^I)\mu_S + t^I \nu_S$. Because $\lambda$ is a convex combination of two systems of balancing weights of $\calC$, $\sum_{S \in \calC', S \ni i} \lambda_S = 1$ for all the players $i \in N$. Concerning player $p$,
\[
\sum_{\substack{S \in \calC' \\ S \ni p}} \lambda_S = \lambda_I = (1-t^I) \mu_I + t^I \nu_I = \mu_I + t^I(\nu_I - \mu_I) = \mu_I + 1 - \mu_I = 1.
\] 
We conclude that $\calC'$ is a balanced collection. Now, let us prove the
minimality of $\calC'$ as a balanced collection. Because ${\rm
    rk}\left( A^\calC \right) = k-1$, the set of systems of balancing weights
for $\calC$ is the set of convex combinations of $\mu$ and $\nu$, and therefore the set of systems of balancing weights for $\calC'$ is a subset of this. More precisely, it is the subset $\{\lambda \in {\rm conv}(\mu, \nu) \mid \lambda_I = 1\}$, equivalently $\{t \in [0, 1] \mid (1-t)\mu_I + t\nu_I = 1\} = T$, and therefore the condition is on the variable $t$. By assumption, $\mu_I \neq \nu_I$, and then $\mu_I < 1 \leq \nu_I$ without loss of generality. Because the map $f: t \mapsto (1-t)\mu_I + t\nu_I$ is linear and $f(0) < 1$ and $f(1) \geq 1$, there is a unique $t^* \in T$ such that $f(t^*) = 1$, then this unique $t^*$ must be $t^I$. 
\endproof

\paragraph{Final algorithm.}
It is now possible to construct, from the set of minimal balanced collections on a set $N$, the set of minimal balanced collections on another set $N' = N \cup \{p\}$ (see Algorithm \ref{construction}). 

\begin{breakablealgorithm}
\caption{AddNewPlayer} \label{construction}
\begin{algorithmic}[1]
\Require A set of minimal balanced collection $\bbB(N)$ on a set $N$
\Ensure A set of minimal balanced collection $\bbB(N')$ on a set $N' = N \cup \{p\}$
\Procedure{AddNewPlayer}{$\bbB(N), p$}
\For{$(\calC^1, \calC^2) \in \bbB(N) \times \bbB(N)$}
\State $\calC \gets \calC^1 \cup \calC^2$ and $k \gets \lvert \calC \rvert$
\If{$\rk(A^\calC) = k-1$}
\For{$I \subseteq [k]$ \textbf{such that} $t^I \in \ ]0, 1[$}
\State \textbf{for} $i \in I$ \textbf{do} add $S_i \cup \{p\}$ with weights $(1-t^I)\mu_{S_i} + \, t^I \nu_{S_i}$ to $\calC'$
\State \textbf{for} $i \not \in I$ \textbf{do} add $S_i$ with weights $(1-t^I)\mu_{S_i} + \, t^I \nu_{S_i}$ to $\calC'$
\State add $\calC'$ to $\bbB(N')$
\EndFor
\EndIf
\EndFor
\For{$\calC \in \bbB_N$}
\State $k \gets \lvert \calC \rvert$
\For{$I \subseteq [k]$ \textbf{such that} $\lambda^\calC_I \leq 1$}
\State $C' \gets \varnothing$
\State \textbf{for} $i \in I$ \textbf{do} add $S_i \cup \{p\}$ with weights $\lambda^\calC_{S_i}$ to $\calC'$
\State \textbf{for} $i \not \in I$ \textbf{do} add $S_i$ with weights $\lambda^\calC_{S_i}$ to $\calC'$
\State \textbf{if} $\lambda^\calC_I < 1$ \textbf{then} add $\{p\}$ with weight $1 - \lambda^\calC_I$ to $\calC'$
\State add $\calC'$ to $\bbB(N')$
\For{$\delta \in [k] \setminus I$ \textbf{such that} $\lambda_{S_\delta} > 1 - \lambda^\calC_I$}
\State $C' \gets \varnothing$
\State \textbf{for} $i \in I \setminus \{\delta\}$ \textbf{do} add $S_i \cup \{p\}$ with weights $\lambda^\calC_{S_i}$ to $\calC'$
\State \textbf{for} $i \not \in I \cup \{\delta\}$ \textbf{do} add $S_i$ with weights $\lambda^\calC_{S_i}$ to $\calC'$
\State add $S_\delta \cup \{p\}$ with weight $1 - \lambda^\calC_I$ to $\calC'$
\State add $S_\delta$ with weight $\lambda^\calC_{S_\delta} + \lambda^\calC_I - 1$ to $\calC'$
\State add $\overline{\calC}$ to $\bbB(N')$
\EndFor
\EndFor
\EndFor
\Return $\bbB(N')$
\EndProcedure
\end{algorithmic}
\end{breakablealgorithm}

\begin{theorem}
The algorithm \Call{AddNewPlayer}{}, which takes as an input the set of all minimal balanced collections on a set $N$, generates all the minimal balanced collections on the set $N' = N \cup \{p\}$. 
\end{theorem}

\proof
Thanks to the four previous lemmas, the algorithm generates only minimal balanced collections on $N'$. It remains to prove that every minimal collection is generated by this algorithm. Let $\calB$ be a minimal balanced collection on $N'$. If the player $p$ is removed from each coalition of $\calB$, the collection is still balanced. Denote by $\calB_{- p}$ this new collection. 
\begin{itemize}
\item If $\{p\} \in \calB$: as $\calB$ is a minimal balanced collection, $\{\bfone^S \mid S \in \calB\}$ is linearly independent (in $\bbR^{N'}$). Hence, because $\{p\} \in \calB$, there does not exist $S \in \calB$ such that $p \in S$ and $S \setminus \{p\} \in \calB$. Therefore, 
\[
\left\{\bfone^{S \setminus \{p\}} \mid S \in \calB, S \neq \{p\} \right\} = \left\{\bfone^T \mid T \in \calB_{-p} \right\}
\] 
is linearly independent in $\bbR^N$. We conclude that the balanced collection $\calB_{-p}$ must be a minimal balanced collection so that $\calB$ is generated by the second case. 
\item If $\{p\} \not \in \calB$ and there exists $S \in \calB$
  such that $p \in S$ and $T \coloneqq S \setminus \{p\} \in \calB$:  then $T
  \in \calB_{-p}$ and for such pairs the weights of $T$ in the balanced
  collection $\calB_{-p}$ must be the sum of the balancing weights of $S$ and
  $T$ in the balanced collection $\calB$. Doing so, the minimality of $\calB$ implies the minimality of $\calB_{-p}$. Then $\calB$ is generated by the third case. 
\item Assume now that there is no singleton $\{p\}$ in $\calB$, and that there is no $S \in \calB$ that satisfies $p \in S$ and $S \setminus \{p\} \in \calB$. 
\begin{itemize}
\item[$\triangleright$] If $\calB_{-p}$ is a minimal balanced collection, $\calB$ is generated by the first case. 
\item[$\triangleright$] Assume now that $\calB_{-p}$ is not a minimal balanced collection. Because $\calB$ is a minimal balanced collection of $k$ coalitions, ${\rm rk}(A^\calB) = k$, and therefore ${\rm rk}(A^{\calB_{-p}}) = k-1$. Consequently, the set of solutions of the following system of inequalities
\begin{equation}
\label{system}
A^{\calB_{-p}} \lambda = \bfone^N, \quad \lambda_i \geq 0, \forall i \in [k]
\end{equation}
is one-dimensional and has the form $\lambda = \lambda^0 + t\lambda^1$, where $\lambda^0$ is a system of balancing weights for $\calB_{-p}$, $t$ is a real number and $\lambda^1$ is a nonzero solution of the homogeneous system
\[
A^{\calB_{-p}}\lambda = 0, \quad \lambda_i \geq 0, \forall i \in [k].
\]
The set of solutions of (\ref{system}) being bounded and one-dimensional, it is a non-degenerate segment $[\alpha, \beta]$ that consists of all the solutions of the above system. Let $U_\alpha = \{i \mid \alpha_i > 0\}$ and $U_\beta = \{i \mid \beta_i > 0\}$. Clearly, $U_\alpha$ and $U_\beta$ are proper subsets of $\{1, \ldots, k\}$ and $U_\alpha \cup U_\beta = \{1, \ldots, k\}$. Denote $\calB^\alpha = \{B_i  \in \calB \mid i \in U_\alpha\}$ and $\calB^\beta = \{B_i \in \calB \mid i \in U_\beta\}$. $\alpha^*$, the restriction of $\alpha$ to $U_\alpha$, is a system of balancing weights for $\calB^\alpha$, and $\beta^*$, the restriction of $\beta$ to $U_\beta$, is a system of balancing weights for $\calB^\beta$. Since $\alpha$ and $\beta$ are extremal solutions of the system (\ref{system}), $\calB^\alpha$ and $\calB^\beta$ must be minimal balanced collections. Then $\calB$ is the union of $\calB^\alpha$ and $\calB^\beta$, and is generated by the fourth case. 
\end{itemize}
\end{itemize}
\endproof

With the procedure \Call{AddNewPlayer}{} used recursively, all the minimal
balanced collections on any fixed set $N$ are generated from the ones on $\{1,
2\}$. This is achieved by the procedure \Call{Peleg}{} (see Algorithm
\ref{computation}).
\begin{breakablealgorithm}
\caption{Minimal balanced collections computation} \label{computation}
\begin{algorithmic}[1]
\Require A number of players $n \geq 3$
\Ensure The set of minimal balanced collections on the set $[n]$
\Procedure{Peleg}{$n$}
\State $\bbB(\{1, 2\}) \gets \{\{\{1, 2\}\}, \{\{1\}, \{2\}\}\}$
\For{$i \in \{3, \ldots, n\}$}
\State $\bbB([i]) \gets$ \Call{AddNewPlayer}{$\bbB([i-1])$, $i$}
\EndFor
\Return $\bbB([n])$
\EndProcedure
\end{algorithmic}
\end{breakablealgorithm}

An example of generation of minimal balanced collections, illustrating the
different cases, is given in Appendix~\ref{app:B}.

\begin{remark}It is possible to adapt Algorithm \ref{construction} to compute the minimal
balanced collections on every set system $\calF\subseteq 2^N$ on which the game
is defined. The only difference for the implementation is to check, when a new
minimal balanced collection is created, that every coalition is a subset of an
element of the set system. If it is not the case, just ignore the newly created
collection and continue the computation.
\end{remark}

\paragraph{Results and performance.}
We implemented the above algorithms in Python\footnote{Computing device: Intel Xeon W-1250, CPU 3.30 GHz, 32 GB RAM}, and found the following results and
performance, given in Table~\ref{tab:mbc}.
\begin{table}[htb]
\begin{center}
\begin{tabular}{|c|r|c|}
\hline
$n$ & Nb of minimal balanced collections & CPU time  \\
\hline
$1$ & $1$ & - \\
$2$ & $2$ & $ 0.00$ sec \\
$3$ & $6$ & $ 0.01$ sec \\
$4$ & $42$ & $ 0.03$ sec \\
$5$ & $1,292$ & $ 1.05$ sec \\
$6$ & $200,214$ & $ 4$ min $4$ sec \\
$7$ & $132,422,036$ & $ 63$ hours \\
\hline
\end{tabular}
\end{center}
\caption{Number of minimal balanced collections as a function of $n$.}
\label{tab:mbc}
\end{table}
We have checked if this sequence of numbers was already known in the OEIS (On Line Encyclopedia of Integer Sequences (Sloane, 1964)). As it was not the case, we added it to the Encyclopedia, and it can be accessed under the number A355042.\footnote{See {\tt https://oeis.org/A355042}.} Moreover, we have stored all minimal balanced collections till $n=7$.\footnote{Available on request from the
  corresponding author.}

Compared to the number of maximal unbalanced collections (see
Table~\ref{tab:muc}), we see that the latter increase much slower than the number of minimal balanced collections. 

\subsection{Vertex enumeration method}\label{sec:vert}
Consider the polytope $W(N)$ defined by
  \begin{equation}\label{eq:polymbc}
W(N) = \left\{\lambda\in\bbR^{2^N\setminus\{\varnothing\}}\mid \sum_{S\in
  2^N\setminus\{\varnothing\}}\lambda_S\bfone^S=\bfone^N, \lambda_S\geq 0, \forall S\in2^N\setminus\{\varnothing\}\right\}
\end{equation}
 It is easy to check that the vertices of $W(N)$ are in bijection with the
 minimal balanced collections on $N$. Indeed, an element of $W(N)$ is a
   vertex if and only if its support is minimal balanced with the corresponding
   balancing weights (see, e.g., \cite[Corollary 3.1.9]{pesu03}). The reason is essentially the following. Consider
 $\lambda$ an element of $W_N$. By definition, $\calB\coloneqq \{S\in
 2^N\setminus\{\varnothing\}\mid \lambda_S>0\}$ is a balanced collection with
 balancing weight system $\lambda$. If $\lambda$ is a vertex, it cannot be
 obtained as a convex combination of other vectors in $W(N)$, hence the
 balancing weight system is unique and the corresponding balanced
 collection is minimal.

Consequently, generating all minimal balanced collections of $N$ amounts to
finding all vertices of $W(N)$. As described in
  \cite{boelguma09}, vertex enumeration of a polytope remains one of the open
  problem in geometry. We have used the classical Avis-Fukuda (1992) method
for enumerating all vertices, available in the pycddlib package in
Python. Running the algorithm for $n=6$ gave the following (the performance of
our implementation of Peleg's method is recalled), see
Table~\ref{tab:1}.\footnote{Computing device: Apple M1 chip, CPU 3.2 GHz, 16 GB
  RAM.} The comparison clearly shows that Peleg's method outperforms the
Avis-Fukuda algorithm.

\begin{table}[htb]
\begin{center}
\begin{tabular}{|c|c|c|}
\hline
Algorithm used  & Algorithm based on Peleg's method & Avis-Fukuda algorithm \\
\hline
Computation time & $244$ seconds & $1764$ seconds  \\
\hline
\end{tabular}
\caption{Comparison of the computation times of both methods with $n=6$.}
\label{tab:1}
\end{center}
\end{table}
  
\section{Applications in game theory}\label{sec:appl}
This section is devoted to various applications of minimal balanced collections
in cooperative game theory. We will start by recalling the famous problem of
nonemptiness of the core and its classical solution brought by Bondareva and
Shapley independently, which can be viewed as the very starting point of
minimal balanced collections. Then, we will prove some original results, showing
that minimal balanced collections can be used to prove nontrivial properties of
coalitions.

An important general remark for all the subsequent results is that the  set
of minimal balanced collections does not depend on the game
under consideration, but only on $n$. Therefore, there is no need to generate
them at each application, but just to export them from some storage device. Till
$n=7$, this gives a computational advantage compared to other methods based on
linear programming and polyhedra, as it will be shown with the example of the
nonemptiness of the core.  

\subsection{Nonemptiness of the core}\label{sec:none}
Let $(N,v)$ be a game. The question is whether the core $C(N,v)$ of this
game is nonempty. Consider the following linear program:
 \begin{align*}
   \min \ \ & x(N)\\
   \text{s.t. } & x(S)\geq v(S), \forall S\in 2^N\setminus\{\varnothing\}.
 \end{align*}
Clearly, $C(N,v)\neq\varnothing$ if and only if the optimal value of this LP is
$x(N)=v(N)$. Therefore, one simple way to check nonemptiness of the core is to
solve this LP and compute its optimal value.

Another way is to take the dual program of this LP. This was done by Bondareva
\cite{bon63} and Shapley \cite{sha67}, and directly leads to minimal balanced
collections and the following classical result.
  \begin{theorem}\label{th:BS}
A game $(N,v)$ has a nonempty core if and only if for any minimal balanced collection $\calB$ with balancing vector $(\lambda^\calB_S)_{S \in \calB}$, we have 
\begin{equation}\label{eq:none}
\sum_{S \in \calB} \lambda^\calB_S v(S) \leq v(N).
\end{equation}
Moreover, none of the inequalities is redundant, except the one for $\calB = \{N\}$.
\end{theorem}
This result shows that nonemptiness of the core can be checked by a simple
algorithm inspecting inequality (\ref{eq:none}) for each minimal balanced
collection. The test can be stopped once we find a minimal balanced collection
for which the inequality is violated.

\medskip

In order to compare both approaches, we fixed $n=6$ and generated 5000 different
games in the following way: the values $v(S)$ for all coalitions but the grand
coalition $N$ are drawn at random in the interval $[0,5]$, while $v(N)$ is fixed
to 50. Doing so, each generated game has a nonempty core, as
  $(50/6,\ldots,50/6)$ is a core element for any generated game. Therefore, in
the algorithm based on the Bondareva-Shapley theorem, all inequalities have to
be checked in order to conclude for nonemptiness of the core (most defavorable
case).  For solving the LP, we have used the revised simplex method, already
implemented natively in Python. Both algorithms being implemented in the
  same language, the comparison is fair.  The results are given in
Table~\ref{tab:2}.\footnote{Computing device: Apple M1 chip, CPU 3.2 GHz, 16 GB RAM.}
\begin{table}[htb]
\begin{center}
\begin{tabular}{|c|c|c|}
\hline
 Algorithm used  & Algorithm based on BS theorem & Revised simplex method \\
\hline
Cumulated computation time & $0.96$ seconds & $24.85$ seconds  \\
\hline
\end{tabular}
\caption{Comparison of the computation time for checking 5000 games with $n=6$,
  for both methods.}
\label{tab:2}
\end{center}
\end{table}

We conclude that the algorithm based on minimal balanced collections (provided
they are generated off line) is much
faster than a direct approach based on linear programming. 

\subsection{Properties of coalitions and collections}\label{sec:prop}
Throughout the section let $(N,v)$ be a {\it balanced} game, i.e., a game
  with a nonempty core, $S$ be a coalition, and
$S^c \coloneqq N\setminus S$. Denote by $H_S$ the hyperplane of the set of preimputations defined by 
\[
H_S = \{x \in X(N,v) \mid x(S) = v(S)\}. 
\]
Denote by $(S, v)$ the subgame on $S$, in which only the subcoalitions of $S$
are considered, and by $(N, v^S)$ the game that may differ from $(N,v)$ only
inasmuch as $v^S(S^c) = v(N) - v(S)$. This definition can be extended to a
collection of coalitions $\calS$, with $v^\calS(S^c) = v(N) - v(S)$ for all $S
\in \calS$ and $v^\calS(T) = v(T)$ otherwise.

All algorithms pertaining to this section are relegated in Appendix~\ref{app:C}.

\paragraph{Exactness.}
A coalition $S$ is \emph{exact} (for $(N,v)$) if there exists a core element $x \in C(N,v)$ such that $x(S) = v(S)$.

Hence, a coalition $S$ is exact if and only if the hyperplane $H_S$ intersects the core. The following result permits us to build an algorithm that checks exactness. 

\begin{proposition}
\label{exact}
Let $(N,v)$ be a balanced game. A coalition $S$ is exact if and only if $(N, v^S)$ is balanced. 
\end{proposition}

\proof
Assume that $(N,v^S)$ is balanced. Then, for all $x \in C(N,v^S)$, $x(N) = v(N)$ and $x(S^c) \geq v(N) - v(S)$. It implies that $x(S) = x(N) - x(S^c) \leq v(S)$. But, because $x$ belongs to the core of $(N,v^S)$, it follows that $x(S) \geq v(S)$, and therefore $x(S) = v(S)$. \\
Assume now that $S$ is exact. Therefore, there exists $x \in C(N,v)$ such that $x(S) = v(S)$. Because $x(N) = v(N)$, then $x(S^c) = v(N) - v(S)$, and $x \in C(N,v^S)$.
\endproof

\paragraph{Effectiveness.} A coalition $S$ is {\it effective} (for  $(N,v)$) if $x(S)=v(S)$
for all core elements $x$. We denote by $\calE(N,v)$ the set of coalitions that
are effective for $(N,v)$.

Equivalently, $S \in \calE(N,v)$ if and only if $C(N,v) \subseteq H_S$.
The following result allows to obtain all effective coalitions of a game. 

\begin{lemma}
\label{union}
$\calE(N,v)$ is the union of all minimal balanced collections $\calB$ such that 
\[
\sum_{S \in \calB} \lambda^\calB_S v(S) = v(N).
\]
\end{lemma}

\proof Let $\calB$ be a minimal balanced collection such that $\sum_{S \in \calB} \lambda^\calB_S v(S) = v(N)$ and $x$ be a core element. Then
\[
v(N) = x(N) = \sum_{S \in \calB} \lambda^\calB_S x(S) \geq \sum_{S \in \calB} \lambda^\calB_S v(S) = v(N).
\]
As $\lambda^\calB_S > 0$, $x(S) = v(S)$ for all $S \in \calB$, i.e., $\calB \subseteq \calE(N,v)$. \\
For the other inclusion, let $S \in \calE(N,v)$. As $\{N\}$ is a minimal balanced collection, it may be assumed that $S \neq N$. It remains to show that $S$ is contained in some minimal balanced collection $\calB$ that satisfies $\sum_{S \in \calB} \lambda^\calB_S v(S) = v(N)$. Assume the contrary. Then, by Theorem~\ref{th:BS}, there exists $\varepsilon > 0$ such that the game $(N,v^\varepsilon)$ that differs from $(N,v)$ only inasmuch as $v^\varepsilon(S) = v(S) + \varepsilon$ is still balanced. Hence, for $x \in C(N,v^\varepsilon)$, it follows $x(S) > v(S)$ and $x \in C(N,v)$, then the desired contradiction has been obtained.
\endproof

\paragraph{Strict vital-exactness.}
A coalition $S$ is \emph{strictly vital-exact} (for $(N,v)$) if there exists a
core element $x \in C(N,v)$ such that $x(S) = v(S)$ and $x(T) > v(T)$ for all $T
\in 2^S \setminus \{\varnothing, S\}$ (Grabisch and Sudh\"olter
\cite{grsu20}). Denote by $\calVE(N,v)$ the set of strictly vital-exact
coalitions.

In particular, an exact singleton is strictly vital-exact. 

\begin{remark}
\label{midpoint}
Let $(N,v)$ be a balanced game. Because the core is convex, for any collection $\calS$ of coalitions such that $\calS \cap \calE(N,v)$ is empty, there exists a core element $x^\calS$ such that $x^\calS(S) > v(S)$, for all $S \in \calS$. Indeed, for every coalition $S \in \calS$, there exists a core element $x^S$ such that $x^S(S) > v(S)$ because $S \not \in \calE(N,v)$. Then, by taking the convex midpoint $\frac{1}{\mid \calS \mid} \sum_{S \in \calS} x^S$, the desired $x^\calS$ is defined, and it belongs to the core by convexity. 
\end{remark}

By Remark \ref{midpoint}, we deduce that the minimal (w.r.t. inclusion) coalitions of $\calE(N,v)$ are strictly vital-exact. 
In view of Lemma \ref{union}, the following proposition shows that it is possible to compute the set of strictly vital-exact coalitions. 

\begin{proposition}
A coalition $S$ is strictly vital-exact if and only if it is exact and 
\[
\calE(N,v^S) \setminus \{S\} \subseteq \{R \in 2^N \mid R \cap S^c \neq \varnothing\}.
\]
\end{proposition}

\proof Assume that $S$ is strictly vital-exact. Then there exists $x \in C(N,v)$ such that $x(S) = v(S)$ and $x(T) > v(T)$ for all $T \in 2^S \setminus \{\varnothing, S\}$. Therefore, no coalition $T \in 2^S \setminus \{\varnothing, S\}$ is included in $\calE(N,v^S)$. \\
Conversely, assume that $S$ is exact and $\calE(N,v^S) \setminus \{S\} \subseteq \{R \in 2^N \mid R \cap S^c \neq \varnothing\}$. Thanks to Proposition \ref{exact}, $C(N,v^S)$ is nonempty. The collection $2^S \setminus \{S\}$ does not intersect $\calE(N,v^S)$ by hypothesis. Hence, by Remark \ref{midpoint} there exists an element $x \in C(N, v^S)$ such that $x(T) > v(T)$ for all $T \in 2^S \setminus \{\varnothing, S\}$. 
\endproof

\paragraph{Extendability.}
(Kikuta and Shapley \cite{kish86}) A coalition $S$ is called \emph{extendable}
(w.r.t. $(N,v)$) if, for any $x \in C(S,v)$, there exists $y \in C(N,v)$ such
that $x = y_S$, where $y_S$ is the restriction of $y$ to coordinates in $S$. A
game $(N,v)$ is \emph{extendable} if all coalitions are extendable.

To check whether a coalition is extendable, by convexity of the core, it is
sufficient to check if each vertex of $C(S,v)$ can be extended to a core
element. To this end, the reduced game property of the core is used. Let $S$ be
a coalition, and $z \in \bbR^{S^c}$. Recall that the traditional \emph{reduced
  game} (Davis and Maschler \cite{dama65}) of $(N,v)$ w.r.t. $S$ and $z$,
$(S, v_{S, z})$, is the game defined by
\[ v_{S, z} (T) = \left\{ \begin{array}{ll}
v(N) - z(S^c), & \quad \quad \text{if }T = S, \\
\max_{Q \subseteq S^c} v(T \cup Q) - z(Q), & \quad \quad \text{if } \varnothing \neq T \subsetneq S.
\end{array} \right. \]
According to Peleg \cite{pel86} the core satisfies the \emph{reduced game property}, i.e., if $x \in C(N,v)$, then $x_S \in C(S, v_{S, x_{S^c}})$. 

\begin{lemma}
\label{ext}
Let $(N,v)$ be a balanced game, $S$ be a coalition and $y \in C(S,v)$. Then
there exists $x \in C(N,v)$ such that $x_S = y$ if and only if $(S^c, v_{S^c,
  y})$ is balanced.
\end{lemma}

\proof The only if part is due to the reduced game property. For the if part
choose an arbitrary $z \in C(S^c, v_{S^c, y})$. It suffices to show that the
only allocation $x \in \bbR^N$ such that $x_S = y$ and $x_{S^c} = z$ belongs to
the core. Assume, on the contrary, that $x \not \in C(N,v)$. As $x(S^c) =
v_{S^c, y}(S^c) = v(N) - x(S)$ by definition, $x(N) = v(N)$.  Therefore, there
exists $T \subsetneq N$ such that $x(T) < v(T)$. As $(N,v)$ is balanced, $v(S^c)
\leq v(N) - v(S) = v(N) - x(S)$, so that $T \neq S^c$. Moreover, as $y \in
C(S,v)$, $T \cap S^c \neq \varnothing$.  Therefore, $v_{S^c, y}(T \cap S^c) =
\max_{Q \subseteq S} v((T \cap S^c) \cap Q) - x(Q) \geq v((T \cap S^c) \cup (T
\cap S)) - x(T \cap S)$. Hence, $x(T \cap S^c) < v(T) - x(T \cap S) \leq v_{S^c,
  y} ( T \cap S^c)$, which contradicts $x_{S^c} = z \in C(S^c, v_{S^c, y})$.
\endproof

Lemma \ref{ext} gives us a necessary and sufficient condition for the existence
of an extension of an element of $C(S,v)$ to an element of $C(N,v)$, based upon
a balancedness check. If there exists an extension for each extreme point of
$C(S,v)$, by convexity of the core, any element of $C(S,v)$ can be
extended. Algorithm~\ref{extendable} in Appendix~\ref{app:C} checks for each extreme point whether the reduced game of
$(N,v)$ w.r.t. the complement of $S$ and the currently considered extreme point
is balanced.

\paragraph{Feasible collections.}
Let $(N,v)$ be a balanced game and $\calF \subseteq 2^N$ be a core-describing collection of coalitions, i.e.,
\[
C(N,v) = \{x \in X(N,v) \mid x(S) \geq v(S), \forall S \in \calF\}. 
\]
Let us consider a subcollection $\calS\subseteq \calF$, and consider the
following subset of $X(N,v)$:
\[
X_\calS = X_\calS^\calF = \left\{x \in X(N,v) \left\vert \ \begin{array}{l} x(S) < v(S) \text{ for all } S \in \calS, \\ x(T) \geq v(T) \text{ for all } T \in \calF \setminus \calS \end{array} \right. \right\}. 
\]
We may call $X_\calS$ a {\it region} of $X(N,v)$, remarking that in the
hyperplane $X(N,v)$, the hyperplanes $H_S$, $S\in\calF$, form a hyperplane
arrangement, inducing (elementary) regions (see Section~\ref{sec:bcother}). The collection
$\calS$ is $\calF$-\emph{feasible} if the corresponding region $X_\calS^\calF$
is nonempty. The regions form a partition of $X(N,v)$, with $C(N,v) =
X_{\{\varnothing\}}$. If no ambiguity occurs, the collection is simply said to be
feasible, and the region is simply denoted by $X_\calS$. Here are some
properties about the feasible collections.
\begin{lemma}[Grabisch and Sudh\"olter \cite{grsu20}]
\label{block}
Let $(N,v)$ be a balanced game and let $\calS \subseteq \calF$. The following holds. 
\begin{enumerate}
\item[(i)] If $\calS$ is feasible, then it does not contain a balanced collection. 
\item[(ii)] For $S, S' \in \calS$ such that $S \cup S' = N$, no $x \in X_\calS$ is dominated via $S$ or $S'$. 
\end{enumerate}
\end{lemma}
Interestingly, (i) shows that a feasible $\calS$ must be unbalanced (see
Section~\ref{sec:bcother} for a similar result). 

A collection $\calS$ that contains only two coalitions satisfying condition $(ii)$ above is called a \emph{blocking feasible collection}. A characterization that can be translated into an algorithm is needed to compute the set of feasible collections. In the sequel, denote $\calS^c = \{S^c \mid S \in \calS\}$.

\begin{lemma}
A collection $\calS \subseteq \calF$ is feasible (w.r.t. $\calF$) if and only if for every minimal balanced collection $\calB \subseteq \calF' = (\calF \setminus \calS) \cup \calS^c$, 
\begin{equation} 
\label{feasibleeq}
\sum_{T \in \calB} \lambda_T^\calB v^\calS(T) 
\left\{ \begin{array}{ll}
\leq v(N), & \\
< v(N), & \text{ if } \calB \cap \calS^c \neq \varnothing.
\end{array} \right.
\end{equation}
\end{lemma}

\proof For $\varepsilon, \alpha \in \bbR$ define $(N,v^\calS_{\varepsilon, \alpha})$ by, for all coalitions $T$, 
\[
v^\calS_{\varepsilon, \alpha} (T) = \left\{ \begin{array}{ll}
v^\calS(T) + \varepsilon & \quad \quad \text{if } T \in \calS^c, \\
v(T) & \quad \quad \text{if } T \in \calF \setminus (\calS \cup \calS^c) \text{ or } T = N, \\
\alpha & \quad \quad \text{otherwise}.
\end{array} \right.
\]
A collection $\calS$ is feasible if and only if there exists $x \in \bbR^N$ and
$\varepsilon > 0$ such that $x(S) \geq v(S)$ for all $S \in \calF \setminus
\calS$, $x(N) = v(N)$, and $x(P) \leq v(P) - \varepsilon$, i.e., $x(N \setminus
P) = x(N) - x(P) = v(N) - x(P) \geq v(N) - v(P) + \varepsilon$ for all $P \in
\calS$. Therefore, for $\alpha \leq \min_{R \in 2^N} x(R)$, $x \in
C(N,v^\calS_{\varepsilon, \alpha})$ so that if part of the proof is finished by
Theorem~\ref{th:BS}. \\ For the only if part we again employ
Theorem~\ref{th:BS}. Indeed, by (\ref{feasibleeq}), there exist $\varepsilon >
0$ and $\alpha \in \bbR$ small enough such that $(N, v^\calS_{\varepsilon,
  \alpha})$ is balanced. The existence of a core element of $(N,
v^\calS_{\varepsilon, \alpha})$ guarantees that $\calS$ is feasible.  \endproof

\section{Generalization: Minimal balanced sets}\label{sec:mbs}
There is a straightforward generalization of balanced collections on a finite
set $N$ of $n$ elements, obtained by viewing subsets of $N$ through their characteristic
vectors in $\{0,1\}^N$. Indeed, it suffices to replace vectors in $\{0,1\}^N$ by
vectors in $\bbR_+^N$. This leads to the following definition.
\begin{definition}
Let $Z \subseteq \bbR_+^N \setminus \{0\}$ be a finite set. $Z$ is a {\it
  balanced set} if there exists a system $(\delta_z)_{z \in Z}$ of positive weights (called \emph{balancing weights}) such that 
\[
\sum_{z \in Z} \delta_z z = \bfone^N.
\] 
\end{definition}
This notion has been introduced by Grabisch and Sudh\"olter in \cite{grsu20}, as
it appears to play a central role in the study of the stability of the core (see Section~\ref{sec:cost}). 

The classical notions and elementary results for balanced collections straightforwardly extend to balanced sets. A balanced set is \emph{minimal} if it does not contain a proper subset that is balanced. Note that a balanced set is minimal if and only if it has a unique system of balancing weights.  Also, a minimal balanced set must be linearly independent. Hence, it contains at most $n$ elements. 

\medskip 

The above observations lead to a direct method for checking if a finite set $Z=\{z^1,\ldots,z^q\}\subseteq \bbR^N_+$ of at most $n$ elements is a minimal balanced set. Consider the matrix $A^Z$ made by the column vectors $z^1,\ldots,z^q$. Then $Z$ is a minimal balanced set if the following linear system 
\begin{equation}\label{eq:mbs1}
A^Z\delta = \bfone^N
\end{equation}
has a unique solution $\delta\in\bbR^q$  which is positive. By standard results in linear algebra, the existence of a unique solution amounts to check that $\rk(A^Z)=q=\rk[A^Z \ \bfone^N]$.

\medskip

An interesting question is how minimal balanced sets are related to minimal
balanced collections, or how to make one from the other, by either replacing in
a vector $z\in Z$ all nonzero coordinates by 1, or conversely by replacing in a
vector $\bfone^S$, $S$ belonging to a minimal balanced collection, coordinates
equal to 1 by a some nonzero numbers. We will show that none of these
transformations is always successful.

One direction is trivial: consider $n=3$ and
$Z=\{(1,\ 0.4, \ 0), (0, \ 0.6, \ 1)\}$. Then $Z$ is a minimal balanced set, but
the corresponding collection is
$\{\{1,2\},\{2,3\}\}$, which is not balanced. The other direction is more
interesting: Let $\calB$ be a minimal balanced collection and consider
$S'\in\calB$. Build $Z=\{\bfone^S,S\in\calB\setminus\{S'\}\}\cup \{z^{S'}\}$,
with $z^{S'}\in\bbR^N_+$ such
that $z^{S'}\neq \bfone^{S'}$ and $z^{S'}_i> 0$ iff $i\in S'$.  Under which
conditions on $z^{S'}$ is $Z$ a minimal balanced set?

First of all, $Z$ must be a linearly independent set. Let us write the
decomposition in direct sum, using previous notation:
\[
\bbR^N= \Ima(A^{\calB\setminus\{S'\}})\oplus \Ker(A^{\calB\setminus\{S'\}})
\]
where Im and Ker denote the image and kernel of the considered linear mappings. 
Let $\{y^1,\ldots,y^{n-q+1}\}$ be a basis of
$\Ker(A^{\calB\setminus\{S'\}})$, where $q=|\calB|$. Then $z^{S'}$ is
linearly independent of $\{\bfone^S,S\in\calB\setminus\{S'\}\}$ iff
\begin{equation}\label{eq:lini}
z^{S'}\cdot y^i\neq 0, \quad \text{ for some } i\in\{1,\ldots,n-q+1\},
\end{equation}
where $\cdot$ indicates the scalar product.
Now, we have the following result.
\begin{proposition}\label{prop:mbs}
Let $\calB$ be a minimal balanced collection on $N$. Consider a set
$Z=\{\bfone^S,S\in\calB\setminus\{S'\}\}\cup \{z^{S'}\}$, with $\bfone^{S'}\neq z^{S'}\in\bbR^N_+$ and $z^{S'}_i>0$ iff $i\in S'$. A necessary condition for $Z$
to be a minimal balanced set is that $z^{S'}$ satisfies (\ref{eq:lini}) and $z^{S'}\in\Ima(A^\calB)$.
\end{proposition}
\proof Consider $Z$ as above, and the matrices
$A^\calB$ and $A^Z$.  $Z$ is a minimal balanced set iff the
system
\[
A^Z\delta=\bfone^N
\]
has a unique solution which is positive for each coordinate. The system has a
unique solution iff $\rk[A^Z\ \bfone^N] = \rk(A^Z)$, i.e., $\bfone^N$ is in the
span of $Z$ and $A^Z$ has full rank $q := |Z|=|\calB|$. As the condition
(\ref{eq:lini}) is satisfied, $Z$ is an independent set, hence $A^Z$ has full
rank. Moreover, $\calB$ is minimal, hence $A^\calB$ has full rank $q$. Now, as $\calB$
is balanced, $\bfone^N$ is in the span of $\{1^S , S \in\calB\}$, and by linear independence of $Z$ and
the assumption that $z^{S'}$ is in the span of $\{1^S , S \in\calB\}$, the vectors in $Z$ and $\{1^S , S \in\calB\}$ span the same space.
\endproof
Unfortunately, the condition fails to be sufficient as nothing ensures the positivity of the
solution. This is shown by the following example:

\begin{example} Take $n = 4$, and consider $\calB = \{\{1,2,3\}, \{1,2,4\},
  \{1,3,4\}, \{2,3,4\}\}$ and
  \[
Z = \{(1,\ 1,\ 1,\ 0), (1,\ 1,\ 0,\ 1), (1,\ 0,\ 0.1,\ 1), (0,\ 0.2,\ 0.1,\ 0.5)\}.
\]
Then the unique solution $\delta$ of (\ref{eq:mbs1}) is
\[
\delta = (25/31,\ -4/31,\ 10/31,\ 50/31).
\]
\end{example}
Nevertheless, the necessary condition of Proposition~\ref{prop:mbs} can be
useful to discard potential candidates for being a minimal balanced set, as shown
in the next example.
\begin{example}
Take $n=5$, the minimal balanced collection
\[
\calB=\{\{3,4,5\},\{1,2,4,5\},\{2,3\},\{1,3\}\}
\]
and choose $S'=\{1,2,4,5\}$,
letting $z^{S'}=(\alpha,\ \beta, \ 0, \ \gamma, \ \delta)$, with
$\alpha,\beta,\gamma,\delta>0$. We obtain that any vector
$y\in\Ker(A^{\calB\setminus\{\{1,2,4,5\}\}})$ has the form
\[
y=(-\theta,\ -\theta, \ \theta, \ \theta-\rho, \ \rho), \quad \theta,\rho\in \bbR,
\]
hence $(-1,\ -1,\ 1,\ 1,\ 0)$ and $(0,\ 0,\ 0,\ -1,\ 1)$ form a basis of
$\Ker(A^{\calB\setminus\{\{1,2,4,5\}\}})$. Then, condition (\ref{eq:lini})
reads:
\[
\gamma\neq \alpha+\beta \text{ or } \gamma\neq \delta.
\]
Let us examine now the second condition. A vector
$y\in\Ima(A^{\calB\setminus\{\{1,2,4,5\}\}})$ has the form
\[
y=(\lambda_2+\lambda_4,\ \lambda_2+\lambda_3,
\ \lambda_1+\lambda_3+\lambda_4,\ \lambda_1+\lambda_2, \ \lambda_1+\lambda_2),
\]
with $\lambda_1,\lambda_2,\lambda_3,\lambda_4\in\bbR$, 
showing that the two last coordinates must be equal. Combining both conditions,
we find that the necessary condition of Proposition~\ref{prop:mbs} becomes
\[
\gamma\neq \alpha+\beta, \quad \gamma=\delta.
\]
\end{example}

\medskip

As for minimal balanced collections, an enumeration problem arises if one
reduces the choice of a possible $Z$ as a subset of some finite set
$\Omega\subseteq \bbR^N_+$ (for minimal balanced collections, $\Omega=\{1^S,S\in
2^N\setminus\{\varnothing\}\}$). This will be indeed the case when studying the
stability of the core (see Section~\ref{sec:cost}).

Unfortunately, the basic inductive mechanism of Peleg's algorithm does not seem
to be adaptable to this general framework. Then we are left with two methods:
either check by the direct method (see Equation~(\ref{eq:mbs1})) if every subset
of $\Omega$ of at most $n$ elements is a minimal balanced set, possibly taking
advantage of Proposition~\ref{prop:mbs}) to eliminate candidates, or use the
polyhedral approach. Indeed, it is
easy to see that minimal balanced sets of $\Omega$ are in one-to-one
correspondence with the vertices of the following polytope:
\[
W(\Omega) = \left\{\delta \in \bbR^\Omega \mid \sum_{z \in \Omega} \delta_z z = \bfone^N, \delta_z
\geq 0, \forall z \in \Omega\right\}
\]
which is defined very similarly to (\ref{eq:polymbc}).
Therefore, finding the minimal balanced sets is equivalent to a
  vertex enumeration problem, which is a fundamental open problem in
  geometry. For more details about vertex enumeration problems of polytopes and
  polyhedra defined by a set of inequalities and related problems, we refer the
  reader to \cite{boelguma09}.

\section{Core stability by nested balancedness}\label{sec:cost}
Checking the stability of the core of a game constitutes a nice and complex
application of both minimal balanced collections and minimal balanced sets. The
method is based on a result by Grabisch and Sudh\"olter \cite{grsu20}, giving a
sufficient and necessary condition for core stability. The aim
  of this section is to focus on the
  algorithmic aspects of the result, and the actual implementation as a
  computer program. The result of \cite{grsu20} provides a theoretical
  characterization of core stability, but does not provide a way to compute all
  the intermediary objects involved in it, such as minimal balanced collections,
  minimal balanced sets, nor a way to find exact coalitions, strictly vital exact coalitions,
  feasible collections of coalitions, etc.  The results we have established above
  permit to find them solely by means of the minimal balanced
  collections, thus providing faster algorithms than usual linear programming
  methods, so that finally we come up with a complete
  algorithm able to check core stability.  Moreover, we
  complete the algorithmic characterization of core stability by adding some
  checking procedures of necessary or sufficient conditions for core stability
  at an early stage of the algorithm.

\medskip

Essentially, the method of Grabisch and Sudh{\"o}lter uses a double balancedness condition, called ``nested balancedness''.  We briefly explain why such a condition arises.

Recall that testing nonemptiness of the core is a problem involving one
quantifier on a variable in an uncountable set, and linear inequalities:
\[
\exists x\in X(N,v), x(S)\geq v(S),\forall S.
\]
The Bondareva-Shapley theorem permits to reduce this problem to a finite number
of tests, by using minimal balanced collections. On the other hand, checking core
stability involves two quantifiers on two variables in uncountale sets, and
linear inequalities:
\[
\forall y\in X(N,v)\setminus C(N,v),\exists x\in C(N,v), \exists S\in
2^N,x_i>y_i,\forall i\in S,x(S)=v(S).
\]
Intuitively, one can get rid of each quantifier by a balancedness condition, and
therefore to have {\it in fine} a finite number of tests. As
it will be explained, the second balancedness condition will involve minimal
balanced sets. 

Throughout this section, $(N,v)$ is a balanced game. All definitions and results
in Sections~\ref{sec:aao} to \ref{sec:fin} are due to Grabisch and Sudh\"olter
\cite{grsu20}. 

\subsection{Simple conditions for stability}
We start with giving some necessary or sufficient conditions for core stability, yielding to
simple tests (all involving previously introduced notions and checkable by some
balancedness conditions) permitting to give a quick answer without invoking the
full method.

\begin{proposition}\label{gillies}(Gillies \cite{gil59})
A balanced game has a stable core only if each singleton is exact. 
\end{proposition}
As the set of exact coalitions can be computed (see Section~\ref{sec:prop}), the
necessary condition of Gillies can be easily checked. Another interesting consequence
of this result is the expansion of the space in which the core is externally
stable if it is a stable set. Indeed, if a balanced game satisfies this
necessary condition, for all player $i \in N$, the core element $x \in C(N,v)$
such that $x_i = v(\{i\})$ dominates every element $y$ of $X(N,v)$ such that
$y_i < v(\{i\})$, via $\{i\}$. Therefore, in the definition of stability (see
Sect. \ref{sec:stab}) for the core, the set $I(N,v)$ may be replaced by
$X(N,v)$. 

Recall that $\calVE(N,v)$ denotes the set of strictly vital-exact
  coalitions (see Section~\ref{sec:prop}).
\begin{proposition}
\label{describing}
Let $(N,v)$ be a balanced game. The core is stable only if $\calVE(N,v)$ is core-describing, i.e.,
\[
C(N,v) = \{x \in X(N,v) \mid x(S) \geq v(S), \forall S \in \calVE(N,v)\}.
\]
\end{proposition}

\proof Assume that the core is stable, and  suppose by contradiction that
  there exists $y\in X(N,v)\setminus C(N,v)$ such that $y(S)\geq v(S)$ for all
  $S \in \calVE(N,v)$. Because the core is stable, there exists $x \in C(N,v)$
such that $x \dom y$. Choose a minimal (w.r.t. inclusion) coalition $S$ such
that $x \domS y$. Then, $v(S)=x(S)>y(S)$, and  $x(T) > v(T)$ for all $T
\in 2^S \setminus \{\varnothing, S\}$. Therefore, $S$ is strictly vital-exact,
    a contradiction.  \endproof

This
important result shows that checking core stability should begin by finding all
strictly vital-exact coalitions (using Algorithm~\ref{sve}), and to check if
these coalitions determine the core. If yes, one should work on
$\calF=\calVE(N,v)$ instead of $2^N$, as this considerably reduces the
combinatorial aspect of the method. Hence, for the rest of
Section~\ref{sec:cost}, we put $\calF=\calVE(N,v)$.

\medskip

Kikuta and Shapley \cite{kish86} have provided a sufficient condition for a game
to have a stable core via extendability. 
\begin{theorem}[Kikuta and Shapley \cite{kish86}]
An extendable game has a nonempty and stable core.
\end{theorem}
Extendability of a game can be checked by using Algorithm~\ref{extendable}, but
is time-consuming. However, this property can be considerably weakened as
follows. Say that a game $(N,v)$ is {\it $\calF$-weakly extendable} if each
$\calF$-feasible collection of coalitions (see Section~\ref{sec:prop}) of
$\calF$ contains a minimal (w.r.t. inclusion) coalition that is extendable.
\begin{proposition}\label{prop:ext}
A $\calF$-weakly extendable game  has a nonempty and stable core.
\end{proposition}
\proof Let $\calS$ be a $\calF$-feasible collection and $S$ be extendable and
minimal w.r.t. inclusion in $\calS$. Take $y\in X_\calS$. Then $y(S)<v(S)$ and
$y(T)\geq v(T)$ for all $T\subset S$, $T\in\calF$. Define $z_S\in \bbR^S$ by
$z_S=y_S+\frac{1}{|S|}(v(S)-y(S))\bfone^S$. Clearly, $z_S\in C(S,v)$ and
$(z_S)_i>y_i$ for all $i\in S$. As $S$ is extendable, there exists $x\in C(N,v)$
such that $x_S=z_S$. Then $x\domS y$.  \endproof

\subsection{Association, admissibility, and outvoting}\label{sec:aao}
We first recall the definition of \emph{outvoting},  a transitive subrelation of domination, that was inspired by a definition given by Kulakovskaja (1973). In view of Proposition \ref{describing}, throughout we assume that $(N,v)$ is a balanced game for which the collection of strictly vital-exact coalitions is core-describing.  %Let $\calF$ be a core-describing collection of exact coalitions that contains the set of strictly vital-exact coalitions. 

\begin{definition}
A preimputation $y$ \emph{outvotes} another preimputation $x$ via $P \in \calF$, written $y \succ_P x$, if $y \ {\rm dom}_P \ x$ and $y(S) \geq v(S)$ for all $S \not \in 2^P$. Also, $y$ outvotes $x$, ($y \succ x$) if there exists a coalition $P \in \calF$ such that $y \succ_P x$. 
\end{definition}

Denote by $M(v) = \{x \in X(N,v) \mid y \not \succ x, \forall y \in X(N,v)\}$ the set of preimputations that are maximal w.r.t. outvoting. 

\begin{proposition}
Let $(N, v)$ be a balanced game. Then $C(N,v) = M(v)$ if and only if $C(N,v)$ is stable. 
\end{proposition}

All results are based on this new characterization. To present the main result, some definitions are needed. 

\begin{definition}
Let $S$ be a strictly vital-exact coalition and $\calB$ be a minimal balanced
collection in $\calF$. $\calB$ is \emph{associated with} $S$ if there exists $i \in S$ such that $\{i\} \in \calB$ and 
\[
\calB \subseteq \{\{j\} \mid j \in S\} \cup \{S^c\} \cup (\calF \setminus 2^S).
\]
Denote by $\bbB^S(N)$ the set of minimal balanced collections on $N$ associated with $S$. 
\end{definition}

\begin{example} 
Let $N = \{1, 2, 3, 4\}$, $\calF=\calVE(N,v) = 2^N \setminus \{\emptyset\}$ and $S = \{1, 2\}$. Therefore, the minimal balanced collection $\calB = \{\{1\}, \{2\}, \{3, 4\}\}$ is included in $\bbB^S(N)$. Indeed, the coalitions $\{1\}$ and $\{2\}$ are singletons of $S$, and $\{3, 4\}$ is the complement of $S$. Moreover, $\calB$ is also associated with $\{1, 2, 3\}$ for example. 
\end{example}

Let $S$ be a strictly vital-exact coalition and $\calB$ be a minimal balanced collection associated with $S$. Denote by $\calB^*_S$ the collection $\calB^*_S = \calB \setminus \{\{i\} \mid i \in S\}$. Thanks to the notions of association and outvoting, the following result holds. 

\begin{theorem}
Let $x$ be a preimputation, and $S$ a strictly vital-exact coalition. Then $x$ is outvoted by some preimputation via $S$ if and only if 
\begin{equation}
\forall \calB \in \bbB^S(N), \quad \quad \sum_{\substack{i \in S \\ \{i\} \in \calB}} \lambda^\calB_{\{i\}} x_i + \sum_{T \in \calB^*_S} \lambda^\calB_T v^S(T) < v(N).
\end{equation}
\end{theorem}

This result can be sharpened, with the use of the following notion. 

\begin{definition}
Let $\calS$ be a nonempty collection of strictly vital-exact coalitions, $S \in \calS$, and $\calB$ be a minimal balanced collection associated with $S$. $\calB$ is \emph{admissible} for $\calS$ if $\calB^*_S \cap \calS \neq \emptyset$ or $\calB^*_S \cap \calS^c = \emptyset$. Denote by $\bbB^S_\calS(N)$ the set of minimal balanced collections associated with $S$ and admissible for $\calS$. 
\end{definition}

\begin{example}
Let $N = \{1, 2, 3, 4\}$, $\calF=\calVE(N,v) = 2^N \setminus \{\emptyset\}$ and $S = \{1, 4\}$. Therefore, the collection $\calB = \{\{1, 2\}, \{1, 3\}, \{2, 3\}, \{4\}\}$ is associated with $S$, and $\calB^*_S = \calB \setminus \{\{4\}\}$. Let $\calS = \{\{2, 3\}, \{1, 4\}\}$. The first condition of the definition is satisfied: $\calB^*_S \cap \calS = \{\{2, 3\}\} \neq \emptyset$, therefore $\calB$ is admissible for $\calS$. 
\end{example}

For each nonempty collection $\calS$ of strictly vital-exact coalitions, denote 
\[
\bbC(\calS) = \bigtimes_{S \in \calS} \bbB^S_\calS(N).
\] 
The concept of admissibility allows to sharpen the previous result, and then to reduce the algorithmic complexity of the core stability checking, thanks to the following result. 

\begin{corollary}
Let $\calS$ be a feasible collection. $M(v) \cap X_\calS \neq \emptyset$ if and only if there exists a system of balanced collections $(\calB_S)_{S \in \calS} \in \bbC(\calS)$ and $x \in X_\calS$ such that 
\[
\forall S \in \calS, \quad \quad \sum_{\substack{i \in S \\ \{i\} \in \calB}} \lambda^\calB_{\{i\}}x_i + \sum_{T \in \calB^*_S} \lambda^\calB_T v^S(T) \geq v(N). 
\]
\end{corollary}

\subsection{Minimal balanced subsets} \label{mbsubsets}

For the study of core stability, minimal balanced subsets of a specific set must be computed. Let $\calS$ be a feasible collection and $(\calB_S)_{S \in \calS} \in \bbC(\calS)$. For $S \in \calS$, let $z^S \in \bbR^N$ be given by
\[
z^S_j = \left\{ \begin{array}{ll}
\lambda^\calB_{\{i\}}, & \quad \text{if } j = i \text{ for some } i \in S \text{ such that } \{i\} \in \calB_S, \\
0, & \quad \text{for all other } j \in N. 
\end{array} \right.
\]
Define the sets 
\[ \begin{aligned}
\Omega_A(\calS) & = \left\{ \bfone^{S^c} \mid S \in \calS\right\}, \quad
\Omega_B({\calS}) = \left\{ \bfone^T \mid T \in \calF \setminus \calS
\right\}, \\ \Omega_C(\calS, (\calB_S)_{S \in \calS}) & = \left\{ z^S \mid S \in
\calS \right\}, \\ \text{and} \quad {\Omega:=}\Omega(\calS, (\calB_S)_{S
  \in \calS}) & = \Omega_A({\calS}) \cup \Omega_B({\calS}) \cup \Omega_C(\calS, (\calB_S)_{S \in \calS}). 
\end{aligned} \]
Let \Call{LinAlgSolve}{} be a procedure that takes as an input a $(n \times
k)$-matrix $A$ and returns a $k$-dimensional vector $\lambda$ such that $A
\lambda = \bfone^N$. Denote by $\bbB(\Omega)$ the set of minimal balanced
subsets of $\Omega$. Algorithm~\ref{balancedsets} in Appendix~\ref{app:D}
generates $\bbB(\Omega)$.

\subsection{Nested balancedness}\label{sec:fin}
Finally, for each $z \in \Omega$, we define $a_z = a_z(\calS, (\calB_S)_{S \in \calS}) = \max(A \cup B \cup C)$, where
\[ \begin{aligned}
A & = \{v(N) - v(S) \mid S \in \calS, \bfone^{S^c} = z\}, \\
B & = \{v(T) \mid T \in \calF \setminus \calS, \bfone^T = z\}, \\
C & = \{v(N) - \sum_{T \in \calB^*_S} \lambda_T^{\calB_S} v^S(T) \mid S \in \calS, z = z^S\}.
\end{aligned} \]
Note that $A$ and $B$  are empty or singletons, but $C$ can be multi-valued because distinct coalitions can generate the same $z$. Let $N = \{1, 2, 3\}$, $S = \{1, 2\}$, $T = \{1, 3\}$ and $\calB_S = \calB_T = \{\{1\}, \{2, 3\}\}$. Then, $z^S = (1, 0, 0) = z^T$. To summarize,  
\[
a_z = \left\{ \begin{array}{ll}
\max C & \quad \text{if } C \neq \emptyset = A, \\
\max\{A, C\} & \quad \text{if } C \neq \emptyset \neq A, \\
v(N) - v(S) & \quad \text{if } z = \bfone^{S^c} \text{ for some } S \in \calS, C = \emptyset, \\
v(T) & \quad \text{if } z = \bfone^T \text{ for some } T \in \calF \setminus \calS, A = \emptyset = C.
\end{array} \right.
\]
Recall that $\bbB(\Omega)$ is the set of all minimal balanced sets $Z \subseteq \Omega$ and denote by $\bbB_0(\Omega)$ the subset of $\bbB(\Omega)$ such that, for all $Z \in \bbB_0(\Omega)$, there exists $S \in \calS$ such that $z = \bfone^{S^c} \in Z$ and $a_z = v(N) - v(S)$. 

\begin{theorem}[Grabisch and Sudh\"olter, 2021]
\label{final}
A balanced game $(N,v)$ has a stable core if and only if, for every feasible collection $\calS$ and for every $(\calB_S)_{S \in \calS} \in \bbC(\calS)$, 
\[ \begin{aligned}
\exists Z \in \bbB(\Omega) \setminus \bbB_0(\Omega) \text{ such that } \sum_{z \in Z} \lambda_z^Z a_z & > v(N), \text{ or} \\
\exists Z \in \bbB_0(\Omega) \text{ such that } \sum_{z \in Z} \lambda_z^Z a_z & \geq v(N).
\end{aligned} \]
\end{theorem}

For each $Z \in \bbB(\Omega)$, let $\psi(Z) = \sum_{z \in Z}\lambda_z^Z
a_z$. Algorithm~\ref{nestedbalancedness} in Appendix~\ref{app:D} checks whether
a game has a stable core.

\subsection{Complete algorithm}\label{sec:comp}

In this paper, we presented all the subroutines used for the
  construction of the core stability checking algorithm. Many of them have
  interest of their own, but altogether they allow us to check the core
  stability of a given game. The first necessary subroutine is the computation
  of the feasible collections, that discretizes the set of preimputations. The next
  subroutine computes the set $\bbC(\calS)$, for each feasible collection
  $\calS$. Then, the third one computes the set $\Omega$, for each collection
  $(\calB_S)_{S \in \calS} \in \bbC(\calS)$. Once the sets $\Omega$ are computed,
  we compute the set of its minimal balanced subsets. The last step is to
  compute the weighted sums described in Theorem \ref{final}.

In addition to the necessary subroutines, we described
  subroutines to avoid useless computation. First, we check the non-emptiness of
  the core, thanks to the minimal balanced collections. Next, we compute the set
  $\calVE$ of strictly vital-exact coalitions. Indeed, a necessary condition for
  core stability is that $\calVE$ is core determining. Moreover, Theorem
  \ref{final} works for any core-describing collection, and because $\calVE$ is
  the smallest one, we avoid the generation of useless feasible
  collections. With the subroutine checking the exactness of coalitions, we
  check the exactness of the singletons (see Prop. \ref{gillies}). Then, we
  check if there is no blocking feasible collection, i.e., a feasible collection
  $\calS = \{S_1, S_2\}$ such that $S_1 \cup S_2 = N$ (see
  Prop. \ref{block}). Finally, we compute the set of extendable coalitions and
  discard the feasible collections that contain a minimal (w.r.t. inclusion)
  coalition that is extendable. If we discard them all, the core is stable (see
  Prop. \ref{prop:ext}). The complete algorithm is presented below in
  pseudocode (for the subroutines, see the appendices).

\begin{breakablealgorithm}
\caption{Core stability checking algorithm} \label{core-stability}
\begin{algorithmic}[1]
\Require A game $(N,v)$
\Ensure The Boolean value: `$(N,v)$ has a stable core'
\Procedure{IsCoreStable}{$(N,v)$}
\State $\bbB(N) \gets$ \Call{Peleg}{$\lvert N \rvert$}
\For{$(\calB, \lambda^\calB) \in \bbB(N)$} \algorithmiccomment{Checking balancedness}
\If{$\sum_{S \in \calB} \lambda_S v(S) > v(N)$}
\Return \textbf{False}
\EndIf
\For{$i \in N$} \algorithmiccomment{Checking exactness of the singletons}
\If{$\sum_{S \in \calB} \lambda_S v^{\{i\}} > v(N)$}
\Return \textbf{False}
\EndIf
\EndFor
\EndFor
\State $\calVE \gets \emptyset$, $\mathcal{E}xt \gets \emptyset$
\For{$S \in 2^N \setminus \{\emptyset\}$}
\If{\Call{IsStrictlyVitalExact}{$S$, $(N, v)$}}
\State Add $S$ to $\calVE$
\EndIf
\If{\Call{IsExtendable}{$S$, $(N,v)$}}
\State Add $S$ to $\mathcal{E}xt$ 
\EndIf
\EndFor
\If{\textbf{not} \Call{IsCoreDescribing}{$\calVE$}} \algorithmiccomment{see Prop. \ref{describing}}
\Return \textbf{False}
\EndIf
\For{$\calS \subseteq \calVE$}
\If{\Call{IsFeasible}{$\calS$, $\calVE$, $(N,v)$}}
\If{$\calS = \{S_1, S_2\}$ such that $S_1 \cup S_2 = N$} \algorithmiccomment{see Lemma \ref{block}}
\Return \textbf{False}
\Else
\For{$S \in \mathcal{E}xt$} \algorithmiccomment{see Prop. \ref{prop:ext}}
\If{$S$ minimal (w.r.t. inclusion) in $\calS$}
\State Go to the next set of coalitions $\calS$
\EndIf
\EndFor
\State $\bbC(\calS) \gets$ \Call{Admissibles}{$\calS$, $\calF$, $\bbB(N)$}
\If{\textbf{not} \Call{IsGSConditionSatisfied}{$\calS$, $\calF$, $\bbB(N)$, $(N,v)$}}
\Return \textbf{False}
\EndIf
\EndIf
\EndIf
\EndFor
\Return \textbf{True}
\EndProcedure
\end{algorithmic}
\end{breakablealgorithm}

\subsection{Examples}\label{examples}

Computing device: Apple M1 chip, CPU 3.2 GHz, 16 GB RAM.

\paragraph{4-player game}

Let $(N,v)$ be the game defined by $N = \{1, 2, 3, 4\}$ and $v(S) = 0.6$ if $\lvert S \rvert = 3$, $v(N) = 1$ and $v(T) = 0$ otherwise. The algorithm returns that the set $\calE(N, v)$ only contains $N$. The set of strictly vital-exact coalitions is $\calVE(N,v) = \{\{i\} \mid i \in N\} \cup \{N \setminus \{i\} \mid i \in N\}$. The collection $\{\{1, 3, 4\}, \{1, 2, 3\}\}$ is a blocking feasible collection, so by Lemma \ref{block}, the core is not stable. The CPU time for this example is 0.1 second. 

\paragraph{5-player game}

Let $(N,v)$ be the game defined by Biswas et al. (1999), defined on $N = \{1, 2, 3, 4, 5\}$ by $v(S) = \max\{x(S),y(S)\}$ with $x = (2, 1, 0, 0, 0)$ and $y = (0, 0, 1, 1, 1)$. For this game, the set of effective proper coalitions is
\[
\calE(N,v) \setminus \{N\} = \{\{2, 3\}, \{2, 4\}, \{2, 5\}, \{1, 3, 4\}, \{1, 3, 5\}, \{1, 4, 5\}\}.
\]
The set of strictly vital-exact coalitions is $\calVE(N,v) = \calE(N,v) \cup \{\{i\} \mid i \in N\}$. The feasible collections that do not contain a minimal extendable coalition are the nonempty subsets of $\{\{1, 3, 4\}, \{1, 3, 5\}, \{1, 4, 5\}\}$, so there are 7 feasible collections. The collection $\{\{1, 3, 5\}, \{1, 4, 5\}\}$ does not satisfy the condition of Theorem \ref{final}, therefore the core of the game is not stable. The CPU time for this example is 1.5 seconds. \\

Let $(N,v)$ be the same game, but with $v(N) = 3.1$. The set $\calE(N,v)$ becomes $\{N\}$. The set of strictly vital-exact coalitions now contains 14 coalitions, while the previous game had 11 strictly vital-exact coalitions. The additional ones are $\{1, 3\}, \{1, 4\}, \{1, 5\}$. The set of feasible collections that do not contain a minimal extendable coalition considerably increases, with 51 feasible collections, but no blocking feasible collection. The largest feasible collection contains 6 strictly vital-exact coalitions. The estimated time for the algorithm to check if this specific collection satisfies the condition of Theorem \ref{final} is greater than 200 hours, due to the cardinality of the set $\bbC(\calS)$ with $\calS$ denoting the specific collection. 

\paragraph{6-player game}

Let $(N,v)$ be the game defined by Studen\'y and Kratochv\'il (2021), defined on $N = \{1, 2, 3, 4, 5, 6\}$ by 
\[ \begin{array}{ll}
v(S) & = 2 \text{ for } S = \left\{ \begin{array}{l}
\{2, 5\}, \{3, 5\}, \{1, 2, 5\}, \{2, 3, 5\}, \{2, 4, 5\}, \{2, 5, 6\}, \{1, 2, 4, 5\} \\
\{1, 2, 4, 6\}, \{1, 2, 5, 6\}, \{2, 4, 5, 6\} \text{ and } \{1, 2, 4, 5, 6\}, 
\end{array} \right. \\
v(S) & = 3 \text{ for } S = \{3, 4, 5\}, \\
v(S) & = 4 \text{ for } S = \left\{ \begin{array}{l}
\{3, 6\}, \{1, 3, 5\}, \{1, 3, 6\}, \{3, 4, 6\}, \{3, 5, 6\}, \{1, 2, 3, 5\}, \\
\{1, 3, 4, 5\}, \{1, 3, 4, 6\}, \{1, 3, 5, 6\}, \{2, 3, 4, 5\} \text{ and } \{1, 2, 3, 4, 5\}, 
\end{array} \right. \\
v(S) & = 6 \text{ for } S = \left\{ \begin{array}{l} 
\{2, 3, 6\}, \{1, 2, 3, 6\}, \{2, 3, 4, 6\}, \{2, 3, 5, 6\}, \\
\{1, 2, 3, 4, 6\} \text{ and } \{1, 2, 3, 5, 6\}, 
\end{array} \right. \\
v(S) & = 8 \text{ for } S = \{3, 4, 5, 6\}, \{1, 3, 4, 5, 6\}, \{2, 3, 4, 5, 6\}, \\
v(N) & = 10 \text{ and } v(T) = 0 \text{ otherwise}. 
\end{array} \] \ 

The set $\calE(N,v)$ is only $\{N\}$. The set of strictly vital-exact coalitions is
\[
\{\{i\} \mid i \in N\} \cup \{\{2, 5\}, \{3, 6\}, \{1, 3, 5\}, \{2, 3, 6\}, \{1, 2, 4, 6\}, \{2, 3, 4, 5\}, \{3, 4, 5, 6\}\} 
\]
and the feasible collections that do not contain a minimal extendable coalition are the nonempty subsets of $\{\{1, 3, 5\}, \{3, 4, 5, 6\}, \{2, 3, 4, 5\}\}$. The feasible collection $\{\{1, 3, 5\}, \{3, 4, 5, 6\}\}$ does not satisfy the condition of Theorem \ref{final}, therefore the core of the game is not stable. The CPU time for this example is 18 minutes and 12 seconds, among which 43 seconds for computing the set of minimal balanced collections on a set of 6 players. 

\section{Concluding remarks} \label{sec:conc}
We have shown in this paper that minimal balanced collections are a central
notion in cooperative game theory, as well as in other areas of discrete
mathematics, and even in physics. As a balanced collection is merely the expression of
a sharing of one unit of resource among subsets, we believe that many more
applications should be possible. 

Just focusing on the domain of cooperative games, the consequences of our
results appear to be of primary importance for the computability of many notions
like exactness, extendability, etc. Indeed, a blind application of the
definition of these notions leads to difficult problems related to polyhedra,
limiting their practical applicability. Thanks to our results, provided minimal
balanced collections are generated beforehand (which is possible since they do
{\it not} depend on the considered game), these notions can be checked very
easily and quickly, as most of the tests to be done reduce to checking simple
linear inequalities.

Generating minimal balanced collections has also permitted implementing an
algorithm testing core stability. The examples in Section \ref{examples} have
shown that, even if for many cases, the answer can be obtained quickly, there
are instances where the computation time goes beyond tractability, due mostly to
the use of minimal balanced sets. Still,
further research is needed to investigate more on minimal balanced sets in order
to overcome this limitation.

\bigskip

\noindent \textbf{Declaration of interest:} None. \\

\noindent \textbf{Funding:} This research did not receive any specific grant from funding agencies in the public, commercial, or not-for-profit sectors.

\bibliographystyle{plain}

\bibliography{../BIB/fuzzy,../BIB/grabisch,../BIB/general}

\appendix

\section{Maximal unbalanced collections in thermal quantum physics}\label{app:A}
In theoretical physics, \emph{thermal quantum field theory} is a set of methods
to calculate expectation values of physical observables of a \emph{quantum field
  theory} at finite temperature. Quantum field theory is a theoretical framework
that combines classical field theory (for example, Newtonian gravitation or
Maxwell's equations of electromagnetic fields), special relativity, and quantum
mechanics. Quantum field theory treats particles as excited states of their
underlying quantum fields, which are more fundamental than the particles. \\

Key objects of quantum field theory are the \emph{correlators}, also called
\emph{Green functions}, that are used to calculate various \emph{observables},
i.e., self-adjoint operators on the Hilbert \emph{space of states} $\bbH$ that
extract some physical properties from a particular state of the studied
system. These correlators are all encoded in a generating functional, called the
\emph{partition function}, analog to the way a sequence of coefficients in
combinatorics is encoded in a generating function. \\

With the \emph{imaginary time formalism}, the difference between the partition
function in thermal quantum field theory and in zero-temperature quantum field
theory is a thermal weight $e^{-\beta H}$, which is actually the action of a
\emph{time-evolution} operator $e^{-iHT}$, that operates a shift in time of
$-i\beta$. The physicists aim to extract the corresponding correlators from this
new partition function. In the computation, a function $\Phi$ appears, that
takes as an input a set of complex energies $\{z_i\}_{i \in I}$ satisfying
$\sum_{i \in I} z_i = 0$, called the \emph{imaginary Matsubara
  energies}. Physicists are interested in the analytic continuations of $\Phi$,
which exist only where
\[
\sum_{i \in J} z_i \not \in \mathfrak{Re}, \qquad \forall J \subseteq I.
\]
We remark that for a given subset of indices $J \subseteq I$, the set $H_J
\coloneqq \{z \in \bbC^{\lvert I \rvert} \mid \sum_{i \in J} z_i \in
\mathfrak{Re}\}$ is a hyperplane of the energy space. It has been proven (Evans,
\cite{eva92}) that the analytic continuations of $\Phi$, which produce
solutions called \emph{(thermal) generalised retarded functions} (Evans,
\cite{eva94}), and the regions of the hyperplane arrangement $\{H_J\}_{J
  \subseteq I}$, called the \emph{restricted all-subset arrangement} (Billera et
al., \cite{bimomowawi12}), are in bijection.

Thanks to the discussion in Section~\ref{sec:bcother}, we see that the generalised retarded functions, produced with the imaginary time formalism studying $n$-body problems, are then in bijection with the maximal unbalanced collections on $N$.

\section{Example of generation of minimal balanced collections}\label{app:B}
Let $N = \{a, b, c, d\}$ and $N' = N \cup \{e\}$. Let $S_1 = \{a, b\}$, $S_2 = \{a, c\}$, $S_3 = \{a, d\}$ and $S_4 = \{b, c, d\}$. Then $\calC = \{S_1, S_2, S_3, S_4\}$ is a minimal balanced collection with the following system of balancing weights $\lambda = \left(\frac{1}{3}, \frac{1}{3}, \frac{1}{3}, \frac{2}{3}\right)$. 

\paragraph{First case} Remark that the set $I = \{1, 4\}$ satisfies the equation $\lambda_I = 1$. Therefore, a minimal balanced collection can be constructed as follows:
\[
\calC' = \{\{a, b, e\}, \{a, c\}, \{a, d\}, \{b, c, d, e\}\}, \text{ with } \left( \frac{1}{3}, \frac{1}{3}, \frac{1}{3}, \frac{2}{3} \right).
\]

\paragraph{Second case} Let $I = \{4\}$. Then $\lambda_I = \frac{2}{3} < 1$. Therefore,
\[
\calC' = \{\{a, b\}, \{a, c\}, \{a, d\}, \{b, c, d, e\}, \{e\}\}, \text{ with } \left( \frac{1}{3}, \frac{1}{3}, \frac{1}{3}, \frac{2}{3}, \frac{1}{3} \right).
\]
is a minimal balanced collection on $N'$. 

\paragraph{Third case} Let $I = \{1, 2\}$ and $\delta = 4$. Then $\lambda_I = \frac{2}{3}$ and $1 - \lambda_{S_\delta} = \frac{1}{3}$. Therefore, $1 > \lambda_I > 1 - \lambda_{S_\delta}$ and the following minimal balanced collection can be constructed:
\[
\calC' = \{\{a, b, e\}, \{a, c, e\}, \{a, d\}, \{b, c, d\}, \{b, c, d, e\}\}, \text{ with } \left( \frac{1}{3}, \frac{1}{3}, \frac{1}{3}, \frac{1}{3}, \frac{1}{3} \right).
\]

\paragraph{Last case} For the last case, consider another framework. Let $N = \{a, b\}$, and $\calC^1 = \{\{a\}, \{b\}\}$, $\calC^2 = \{\{a, b\}\}$ be the only two minimal balanced collections on $N$. Let $\calC$ be the union $\calC = \{\{a\}, \{b\}, \{a, b\}\}$. 
\[
\mu = (1, 1, 0) \text{ and } \nu = (0, 0, 1).
\]
Observe that
\[
{\rm rk}(A^\calC) = {\rm rk} \left( \begin{bmatrix} 1 & 0 & 1 \\ 0 & 1 & 1 \end{bmatrix} \right) = 2 = k-1.
\]
Finally, let $I = \{1, 2\}$. Then $\mu_I = 2$, $\nu_I = 0$, and
\[
t^I = \frac{1 - \mu_I }{\nu_I - \mu_I} = \frac{1}{2} \in \ ]0, 1[. 
\]
The following collection may therefore be constructed:
\[
\calC' = \{\{a, c\}, \{b, c\}, \{a, b\}\}, \text{ with} 
\]
\[
\begin{array}{rcl}
\lambda^{\calC'}_{\{a, c\}} & = (1-t^I)\mu_{\{a, c\}} + t^I\nu_{\{a, c\}} = \frac{1}{2}\mu_{\{a, c\}} & = \frac{1}{2}, \\
\lambda^{\calC'}_{\{b, c\}} & = (1-t^I)\mu_{\{b, c\}} + t^I\nu_{\{b, c\}} = \frac{1}{2}\mu_{\{b, c\}} & = \frac{1}{2}, \\
\lambda^{\calC'}_{\{a, b\}} & = (1-t^I)\mu_{\{a, b\}} + t^I\nu_{\{a, b\}} = \frac{1}{2}\nu_{\{a, b\}} & = \frac{1}{2}.
\end{array}
\]

\section{Algorithms for checking various properties of coalitions and
  collections}\label{app:C}
These algorithms refer to results given in Section~\ref{sec:prop}.

\begin{breakablealgorithm}
\caption{Exactness checking subroutine} \label{alg:exact}
\begin{algorithmic}[1]
\Require A coalition $S$, a balanced game $(N,v)$, the set $\bbB(N)$
\Ensure The Boolean value: `$S$ is exact'
\Procedure{IsExact}{$S$, $\bbB(N)$, $(N,v)$}
\State Define $v^S$ such that $v^S(T) = v(T)$ for all $T \in 2^N \setminus \{S^c\}$, and $v^S(S^c) = v(N) - v(S)$
\For{$(\calB, \lambda) \in \bbB(N)$}
\If{$\sum_{T \in \calB} \lambda_T v^S(T) > v(N)$}
\Return \textbf{False}
\EndIf
\EndFor 
\Return \textbf{True}
\EndProcedure
\end{algorithmic}
\end{breakablealgorithm}

\begin{breakablealgorithm}
\caption{Strict vital-exactness checking algorithm} \label{sve}
\begin{algorithmic}[1]
\Require A coalition $S$, a balanced game $(N,v)$, the set $\bbB(N)$
\Ensure The Boolean value: `$S$ is strictly vital-exact'
\Procedure{IsStrictlyVitalExact}{$S$,  $\bbB(N)$}, $(N,v)$
\For{$\calB \in \bbB(N)$}
\If{$\sum_{T \in \calB} \lambda^\calB_T v^S(T) > v(N)$}
\Return \textbf{False}
\ElsIf{$\sum_{T \in \calB} \lambda^\calB_T v^S(T) = v(N)$}
\For{$T \in \calB$}
\If{$T \cap S^c = \varnothing$}
\Return \textbf{False}
\EndIf
\EndFor
\EndIf
\EndFor
\Return \textbf{True}
\EndProcedure
\end{algorithmic}
\end{breakablealgorithm}

\begin{breakablealgorithm}
\caption{Extendability checking algorithm} \label{extendable}
\begin{algorithmic}[1]
\Require A coalition $S$, a balanced game $(N,v)$
\Ensure The Boolean value: `$S$ is extendable'
\Procedure{IsExtendable}{$S$, $(N,v)$}
\State $\bbB(S^c) \gets$ \Call{Peleg}{$\lvert S^c \rvert$}
\For{$\xi \in {\rm ext}(C(S,v))$}
\State define the reduced game $v_{S^c, \xi}$
\For{$\calB \in \bbB(S^c)$}
\If{$\sum_{T \in \calB} \lambda^\calB_T v_{S^c, \xi}(T) > v(N) - v(S)$}
\Return \textbf{False}
\EndIf
\EndFor
\EndFor
\Return \textbf{True}
\EndProcedure
\end{algorithmic}
\end{breakablealgorithm}

\newpage

\begin{breakablealgorithm}
\caption{Feasibility checking algorithm} \label{feasible}
\begin{algorithmic}[1]
\Require A balanced game $(N,v)$, its support $\calF$, a set $\calS \subseteq
\calF$, the set $\bbB(N)$
\Ensure The Boolean value: `$\calS$ is feasible'
\Procedure{IsFeasible}{$\calS$, $\calF$, $\bbB(N)$, $(N,v)$}
\For{$\calB \in \bbB(N)$ \textbf{such that} $\calB \subseteq 
    (\calF\setminus \calS)\cup\calS^c$} {\bf do}
\If{$\calB \cap \calS^c \neq \varnothing$ \textbf{and} $\sum_{S \in \calB}\lambda_S^\calB v^\calS(S) \geq v(N)$}
\Return \textbf{False}
\ElsIf{$\calB \cap \calS^c = \varnothing$ \textbf{and} $\sum_{S \in \calB}\lambda_S^\calB v^\calS(S) > v(N)$}
\Return \textbf{False}
\EndIf
\EndFor
\Return \textbf{True}
\EndProcedure
\end{algorithmic}
\end{breakablealgorithm}

\section{Algorithms for checking core stability}\label{app:D}

\begin{breakablealgorithm}
\caption{Association/admissibility subroutine} \label{asso}
\begin{algorithmic}[1]
\Require A set of coalition $\calS$, the set of minimal balanced collections $\bbB(N)$, the set $\calF$
\Ensure The set $\bbC(\calS)$
\Procedure{Admissibles}{$\calS$, $\calF$, $\bbB(N)$}
\For{$S \in \calS$}
\State $\calA(S) \gets \emptyset$
\For{$(\calB, \lambda) \in \bbB(N)$}
\If{$\calB \subseteq \{\{j\} \mid i \in S\} \cup \{S^c\} \cup (\calF \setminus 2^S)$ \textbf{and} $\calB \cap \{\{j\} \mid j \in S\} \neq \emptyset$}
\State $\calB^* \gets \calB \setminus \{\{j\} \mid j \in S\}$
\If{$\calB^* \cap \calS \neq \emptyset$ \textbf{or} $\calB^* \cap \calS^c = \emptyset$}
\State Add $\calB$ to $\calA(S)$
\EndIf
\EndIf
\EndFor
\EndFor
\Return $\bigtimes_{S \in \calS} \calA(S)$
\EndProcedure
\end{algorithmic}
\end{breakablealgorithm}

\begin{breakablealgorithm}
\caption{Omega computation subroutine} \label{omega}
\begin{algorithmic}[1]
\Require A set of coalition $\calS$, a collection $(\calB_S)_{S \in \calS} \in \bbC(\calS)$, the set $\calF$
\Ensure The set $\Omega = \Omega_A \cup \Omega_B \cup \Omega_C$
\Procedure{Omega}{$\calS$, $(\calB_S)_{S \in \calS}$}
\State $\Omega_A \gets \emptyset$, $\Omega_B \gets \emptyset$, $\Omega_C \gets \emptyset$
\For{$S \in \calF$}
\If{\textbf{not} $S \in \calS$}
\State Add $\bfone^S$ to $\Omega_B$
\Else
\State Add $\bfone^{S^c}$ to $\Omega_A$ and $z \gets \vec{o}_{\bbR^N}$
\For{$i \in S$}
\If{$\{i\} \in \calB_S$}
\State $z_j \gets \lambda^\calB_{\{i\}}$
\EndIf
\EndFor
Add $z$ to $\Omega_C$
\EndIf
\EndFor
\Return $(\Omega_A, \Omega_B, \Omega_C)$
\EndProcedure
\end{algorithmic}
\end{breakablealgorithm}

\begin{breakablealgorithm}
\caption{Minimal balanced sets computation algorithm} \label{balancedsets}
\begin{algorithmic}[1]
\Require A set $\Omega=\Omega_A\cup \Omega_B\cup \Omega_C$
\Ensure The sets $\bbB(\Omega)$ and $\bbB_0(\Omega)$
\Procedure{IsMinimalBalanced}{$Z$}
\If{${\rm rk}(A^Z) = {\rm rk}(A^Z_\bfone) = \lvert Z \rvert$}
\State $\lambda \gets$ \Call{LinAlgSolve}{$A^Z$}
\If{$\lambda > 0$}
\Return \textbf{True}
\EndIf
\EndIf
\Return \textbf{False}
\EndProcedure
\Procedure{BalancedSets}{$\Omega$}
\For{$Z \subseteq \Omega$ \textbf{such that} $\lvert Z \rvert \leq n$}
\If{\Call{IsMinimalBalanced}{$Z$}}
\State add $Z$ to $\bbB(\Omega)$
\For{$z \in Z$}
\If{$z \in \Omega_A \setminus \Omega_C$}
\State Add $Z$ to $\bbB_0(\Omega)$
\EndIf
\EndFor
\EndIf
\EndFor
\Return $(\bbB(\Omega),$ $\bbB_0(\Omega)$)
\EndProcedure
\end{algorithmic}
\end{breakablealgorithm}

\begin{breakablealgorithm}
\caption{Nested balancedness checking subroutine} \label{nestedbalancedness}
\begin{algorithmic}[1]
\Require A feasible collection $\calS$, the game $(N,v)$, its support $\calF$, and the set $\bbB(N)$
\Ensure The Boolean value: `$\calS$ satisfies the conditions of Theorem \ref{final}' 
\Procedure{IsGSConditionSatisfied}{$\calS$, $\calF$, $\bbB(N)$, $(N,v)$}
\State $\bbC(\calS) \gets$ \Call{Admissibles}{$\calS$, $\calF$, $\bbB(N)$}
\For{$(\calB_S)_{S \in \calS} \in \bbC(\calS)$}
\State $(\Omega_A, \Omega_B, \Omega_C) \gets$ \Call{Omega}{$\calS$, $(\calB_S)_{S \in \calS}$}
\State $(\bbB(\Omega), \bbB_0(\Omega)) \gets$ \Call{BalancedSets}{$\Omega_A \cup \Omega_B \cup \Omega_C$}
\If{$\displaystyle \max_{Z \in \bbB(\Omega)} \psi(Z) \leq v(N)$ \textbf{and} $\displaystyle \arg \max_{Z \in \bbB(\Omega)} \psi(Z) \not \in \bbB_0(\Omega)$}
\Return \textbf{False}
\EndIf
\If{$\displaystyle \max_{Z \in \bbB(\Omega)} \psi(Z) < v(N)$ \textbf{and} $\displaystyle \arg \max_{Z \in \bbB(\Omega)} \psi(Z) \in \bbB_0(\Omega)$}
\Return \textbf{False}
\EndIf
\EndFor
\Return \textbf{True}
\EndProcedure
\end{algorithmic}
\end{breakablealgorithm}

\end{document}